\documentclass[structabstract]{aa}  

\usepackage{graphicx}
\usepackage{natbib} 
\usepackage{txfonts}
\newcommand{\nada}[1]{{#1}}

\begin{document}

   \title{Cyclic Variability of the Circumstellar Disc of the Be Star $\zeta$ Tau}

   \subtitle{II. Testing the 2D Global Disc Oscillation Model}
   
   \author{A. C. Carciofi \inst{1}
	       \and
	       A. T. Okazaki \inst{2}
	       \and
	       J.-B. le Bouquin \inst{3}
	       \and
   	       S. \v{S}tefl \inst{3}
	       \and
	       Th. Rivinius  \inst{3}
	       \and
	       D. Baade  \inst{4}
	       \and
	       J. E. Bjorkman \inst{5}
	       \and
	       C. A. Hummel \inst{4}
          }

\offprints{A. C. Carciofi, \email{carciofi@usp.br}}

   \institute{Instituto de Astronomia, Geof{\'\i}sica e Ci{\^e}ncias
Atmosf{\'e}ricas, Universidade de S\~ao Paulo, Rua do Mat\~ao 1226,
Cidade Universit\'aria, S\~ao Paulo, SP 05508-900, Brazil\\
              \email{carciofi@usp.br}
    \and
    Faculty of Engineering, Hokkai-Gakuen University, Toyohira-ku, Sapporo 062-8605, Japan
 \\
    \and
    ESO - European Organisation for Astronomical Research in the Southern Hemisphere, Casilla 19001, Santiago 19, Chile \\
   \and ESO - European Organisation for Astronomical Research in the Southern Hemisphere, Karl-Schwarzschild-Str. 2, 85748 Garching bei M\"unchen, Germany
    \and    University
of Toledo,  Department of Physics \& Astronomy, MS111 2801 W.
Bancroft Street Toledo, OH 43606 USA \\
             }

\date{Received: $<$date$>$; accepted: $<$date$>$; \LaTeX ed: \today}  
\authorrunning{Carciofi et al.}
\titlerunning{Testing the 2D Global Disc Oscillation Model}

 \abstract
   {About 2/3 of the Be stars present the so called $V/R$ variations, a phenomenon characterized by the quasi-cyclic variation of the ratio between the violet and red emission peaks of the \ion{H}{I} emission lines. These variations are generally explained by global oscillations in the circumstellar disc forming a one-armed spiral density pattern that precesses around the star with a period of a few years.
   }
   {In this paper we model, in a self-consistent way,  polarimetric, photometric, spectrophotometric and interferometric observations of the classical Be star $\zeta$ Tauri. Our primary goal is to conduct a critical quantitative test of the global oscillation scenario.}
   {
   We have carried out detailed three-dimensional, NLTE radiative transfer calculations using the radiative transfer code HDUST. For the input for the code we have used the most up-to-date research on Be stars to include a physically realistic description for the central star and the circumstellar disc.
   We adopt a rotationally deformed, gravity darkened central star, surrounded by a disc whose unperturbed state is given by a steady-state viscous decretion disc model. We further assume that disc is in vertical hydrostatic equilibrium. 
     }
   { 
   By adopting a viscous decretion disc model for $\zeta$ Tauri and a rigorous solution of the radiative transfer, we have obtained a very good fit of the time-average properties of the disc. This provides strong theoretical evidence that the viscous decretion disc model is the mechanism responsible for disc formation.
   With the global oscillation model we have successfully fitted spatially resolved VLTI/AMBER observations and the temporal  $V/R$ variations of the H$\alpha$ and Br$\gamma$ lines. This result convincingly demonstrates that the oscillation pattern in the disc is a one-armed spiral. 
Possible model shortcomings, as well as suggestions for future improvements, are also discussed.
   }
   {}


   \maketitle
%

\section{Introduction}
\nada{
$\zeta$ Tauri is a northern, nearby Be star that has drawn the attention of generations of observers and theoreticians. Its position on the sky makes it a convenient target for both northern and southern telescopes, and, as a result, this star has a very rich observational history.
$\zeta$ Tau has some distinctive features that make it an important laboratory for testing theoretical ideas about processes associated with the Be phenomenon.
First, it has a circumstellar (CS) disc that has shown little or no secular evolution in the past 18 years or so (\citealt{ste09}, hereafter Paper I; \citealt{riv06}).
Second, it shows a very stable $V/R$\footnote{The $V/R$ term refers to the quasi-periodic variation of the ratio between the violet and red emission peaks of circumstellar emission lines. It is observed in $\zeta$ Tau and numerous other Be stars. See Paper I and references therein for more details.} cycle, with a quasi-period of about 1430 days \citep[Paper I; ][]{ste07}.}
\nada{
In Paper I, a compilation of spectroscopic, spectrophotometric, polarimetric and spectropolarimetric observations of $\zeta$ Tau was presented, covering 3 complete $V/R$ cycles from 1992 to the present. The data was complemented by  the first spectro-interferometric observations of this star, made in December, 2006 with AMBER at VLTI.
The reader is referred to Paper I for details about the observations and initial data analysis.}

In this paper we report a detailed modeling of the available data. 
\nada{We adopt a physical model for the central star and the CS disc that is consistent with the most up-to-date research on Be stars, as described below.
For the central star we adopt a model that includes 
rotational effects such as geometrical deformation \citep{dom03} and the latitudinal dependency of the stellar radiation \citep{tow04}. Rotational effects are very important in determining the local radiation field at a given point in the stellar environment, since they redirect part of the flux toward the polar regions.}
For the unperturbed state of the CS disc we adopt a steady-state, viscous decretion disc model
\citep{lee91, por99, oka01,car06b}.
In this model, the material is ejected from the stellar equatorial surface, drifts outwards owing to viscous effects, and forms a  geometrically thin disc with nearly Keplerian rotation \citep{riv06}.

In order to model both V/R variations of the spectral lines and the VLTI/AMBER observations, we use
the global disc oscillation model of \citet{oka97} and \citet{pap92}.
In this model, a one-armed $m=1$ oscillation mode is superposed on the unperturbed  
state of the disc. The model is based on the fact that
$m=1$ oscillation modes are the only possible global modes in nearly  
Keplerian discs such as Be discs \citep{kat83}.
It was first applied to Be stars by \citet{oka91} and then revised by  
\citet{pap92}.

\nada{
Although the observed characteristics of  the $V/R$ variations were explained qualitatively well by the global oscillation model, it was not fully satisfactory for several reasons.} First, the oscillation period is sensitive to the subtle details of disc structure as well as stellar parameters \citep{fir06}. Given that there are usually several stellar and disc parameters that are only loosely constrained, this high sensitivity means that it is always possible to obtain the same period by different combinations of parameters.
\nada{
For example, as far as the period is concerned, the same effect is produced by decreasing the disc temperature or stellar radius, or by increasing stellar mass or rotation velocity, thus making model predictions ambiguous. 
Since we use a variety of observations to constrain our models in this investigation, we are able to significantly reduce these ambiguities.
Second, although the model predicts oscillation modes that precess in the direction of disc rotation (prograde modes), as suggested by observations \citep[e.g.,][]{tel94}, the modes are less confined to the inner part of the disc than expected from the observations.
This can be a problem for isolated Be stars, for which the disc extends to a large distance, but we expect that this is not so serious for truncated small discs such as those in binary Be stars like $\zeta$ Tau}. As a matter of fact, we will demonstrate in later sections that a less confined mode better reproduces the observed data.


In order to critically test whether the global disc oscillation model and the viscous Keplerian scenario are applicable for the CS disc of $\zeta$ Tau, it is necessary to rigorously solve the coupled problems of the radiative transfer, radiative equilibrium and statistical equilibrium in non-local thermodynamic equilibrium (NLTE) conditions for the complex 3D geometries predicted by the global oscillation scenario. 
\nada{A 3D approach for the radiative transfer is important to account for geometrical effects such as escape of radiation towards the optically thin polar regions, or the shielding of part of the disc by a local density enhancement.} We use, for this purpose, the computer code HDUST described in \citet{car08b,car06a} and \citet{car08a,car04} 

\nada{In Sect.~\ref{model_description} we \nada{explain} our main model assumptions}. In Sect.~\ref{results}, we outline the adopted computational procedure and discuss in details the fitting of the observations reported in Paper I. 
Finally, in Sect.~ \ref{discussion} and \ref{conclusions} we discuss our results and present a summary of our conclusions.


\section{Model Description \label{model_description}}

The computer code HDUST is a fully 3D, non-local thermodynamic equilibrium (NLTE) code designed to solve the coupled problems of radiative transfer, radiative equilibrium and statistical equilibrium for arbitrary gas density and velocity distributions.
The NLTE Monte Carlo simulation performs a full spectral synthesis by emitting photons from a rotationally deformed and gravity darkened star. The star is divided in a number of latitude bins \nada{(typically 100)}, each with its effective temperature, gravity and a spectral shape given by the appropriate Kurucz model atmosphere \citep{kur94}.  After emission by the star, each photon is followed as it travels through the envelope (where it may be scattered, or absorbed and reemitted, many times) until it escapes.  
During a simulation, whenever a photon scatters, it changes direction, Doppler shifts, and becomes partially polarized. Similarly, whenever a photon is absorbed, it is not destroyed; it is reemitted locally with a new frequency and direction determined by the local emissivity of the gas.  HDUST includes both continuum processes and spectral lines in the opacity and
emissivity of the gas.  Since photons are never destroyed (absorption is always followed by reemission of an equal energy photon packet), this procedure automatically enforces radiative equilibrium and conserves flux exactly.

The interaction (absorption) of the photons with the gas provides a direct sampling of all the radiative rates, as well as the heating of the free electrons.  Consequently, an iterative scheme is adopted in which \nada{the rate equations are solved at the end of each iteration} to update the level populations and electron temperature. \nada{The process proceeds} until convergence of all state variables is reached.
\nada{The interested reader is referred to  \citet{car06a} for details of the Monte Carlo NLTE solution.}


\subsection{Geometry \label{sect:geo}}

The geometry of the problem is defined in Fig.~1. \nada{The star rotates counterclockwise with the rotation axis parallel to the $z$ direction, and is located at the origin of the cartesian system $(x,y,z)$}. The CS disc is assumed to \nada{lie} in the equatorial plane ($xy$ plane). 
\nada{Since we are investigating} non-axisymmetric precessing discs, 
we define the $x$-axis so that it precesses together with the disc. 
Therefore, both the $x$ and $y$ directions are time-dependent.

\nada{We} define the $(x^{\prime},y^{\prime},z^{\prime})$ system so that the observer is located at $x^{\prime} = \infty$ and the plane of the sky is parallel to the $y^{\prime}z^{\prime}$ plane. This system is obtained by rotating the $xyz$ system first by $\phi$ degrees around the $z$ axis and then by $90-i$ degrees around the $y^{\prime}$ axis.
The angle $i$ ($0\le i<180\degr$) is the viewing angle and the angle $\phi$ ($0\le \phi<360\degr$)  describes the time-dependent position of the $x$-axis.

To complete the geometrical description of the system \nada{we specify} the angle $\gamma$  that the projection of the rotation axis on the plane of the sky, $z^{\prime}$, makes with respect to celestial north. We adopt the usual convention that $\gamma$ is measured east from north and that $\gamma$ lies in the
interval $-180 < \gamma \le180\degr$. The angles $\gamma$, $\phi$ and $i$ are all free parameters.

\subsection{The Central Star}

The adopted stellar parameters are listed in Table~\ref{table:model_parameters}.
As discussed by \citet{riv06}, determining the properties of the central star is difficult because of the large $V \sin i$ \citep[$320\,\rm km\, s^{-1}$,][]{yan90} and the presence of an optically thick CS disc.
The adopted parameters for the central star, together with the disc parameters (Sect.~\ref{disc_structure}), form a consistent set of parameters in the sense that, as we shall see below, they can reproduce well all the observational properties.
However, it must be emphasized that many stellar parameters are somewhat uncertain. For instance, equally good fits are obtained by varying the physical size of the star or the photospheric temperatures by about 10\%.

Given the large value of $V \sin i$, we adopt a rotationally deformed and gravity darkened central star instead of a simple spherical star. 
\nada{Assuming uniform rotation, the stellar rotation rate is given by a single parameter 
$V/V_\mathrm{crit}$,
i.e., the ratio between the equatorial velocity and the critical speed 
$V_\mathrm{crit}=[2GM/(3R_\mathrm{p})]^{1/2}$, where $R_\mathrm{p}$ is the polar radius.
}
We assume the stellar shape to be a spheroid, which is a reasonable approximation for the shape of a rigidly rotating, sub-critical star \citep{fre05}. Given the stellar rotation rate and the polar radius, the equatorial radius, $R_\mathrm{e}$, is determined from the Roche approximation for the stellar surface equipotentials \citep{fre05}. 
For the gravity darkening we use the standard von Zeipel flux distribution \citep{von24}, according to which $F(\theta) \propto g_\mathrm{eff}(\theta) \propto T_\mathrm{eff}^4(\theta)$, where $g_\mathrm{eff}(\theta)$ and $T_\mathrm{eff}(\theta)$ are the effective gravity and temperature at stellar latitude $\theta$.
The stellar rotation rate was assumed to be $V/V_\mathrm{crit}=0.8$.
For the polar radius we have adopted the value $R_\mathrm{p} = 5.9\,R_{\sun}$, typical for an evolved main sequence star of spectral type B1IV (\citealt{har88}; Paper I). 

The value of $V_\mathrm{crit}$ in Table~\ref{table:model_parameters} was determined from the data fitting in order to best reproduce the emission line profiles (Sect.~\ref{2dmodel}). The value of the stellar mass was then calculated from $V_\mathrm{crit}$ and $R_\mathrm{p}$. Both $M$ and $V_\mathrm{crit}$ \nada{we obtain} agree with the values tabulated by \citet{har88}.

\subsection{Disc Structure \label{disc_structure}}

   \begin{figure}
   \centering
   \includegraphics[width=\columnwidth]{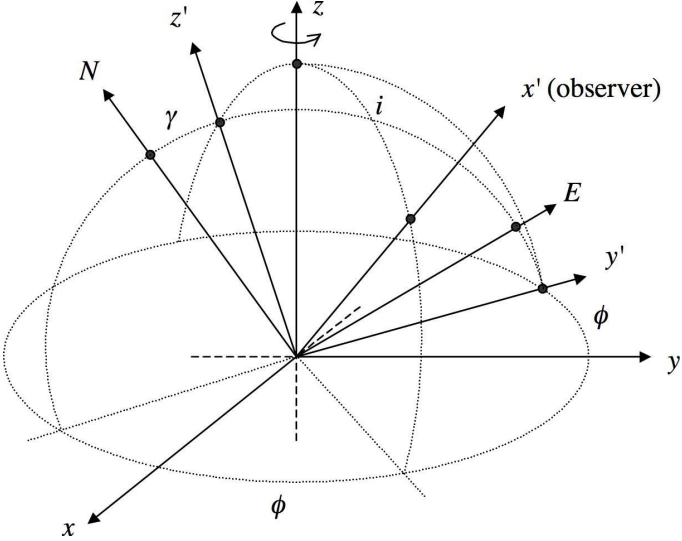}
      \caption{Geometry of the problem. The star is located at the center of the $xyz$ system, with the rotation axis aligned with the $z$ axis. The disc is in the $xy$ plane.
      The observer lies along the $x^{\prime}$ direction, which is defined by the viewing angle $i$ and the azimuthal angle $\phi$, which is the angle between the $x$ axis and the projection of $x^{\prime}$ onto the $xy$ plane.
      The plane of the sky corresponds to the $y^{\prime}z^{\prime}$ plane, and $z^{\prime}$ makes an angle  $\gamma$ with respect to celestial north
              }
         \label{geometry}
   \end{figure}
 %

\begin{table}
\begin{minipage}[t]{\columnwidth}
      \caption[]{Model Parameters for $\zeta$ Tau}
	\label{table:model_parameters}      
	\centering          
	\renewcommand{\footnoterule}{}  
	\begin{tabular}{ l l l l}  
	\hline\hline       
Parameter & Value & Type & Ref.\footnote{References: 1) this work; 2) Paper I; 3) \citet{yan90};
  4) \citet{har88}; 5) \citet{cra05}; 6) \citet{per97}; 7) \citet{har84}}  \\
\hline
\multicolumn{4}{c}{Stellar Parameters}\\ 
\hline
Spectral Type & B1IVe shell & fixed & 2, 3 \\
$R_\mathrm{p}$ & $5.9\,R_{\sun}$  & fixed & 2, 4 \\
$R_\mathrm{e}$ & $7.7\,R_{\sun}$  & fixed\footnote{$R_\mathrm{e}$ was calculated from $R_\mathrm{p}$ and $V/V_\mathrm{crit}$ for a rigidly rotating star in the Roche approximation.} & 1 \\
$T_\mathrm{p}$ & $25\,000\,\rm K$  & fixed &  2 \\
$T_\mathrm{e}$ & $17\,950\,\rm K$  & fixed\footnote{The equatorial temperature, $T_\mathrm{e}$, was calculated from $V/V_\mathrm{crit}$ and $T_\mathrm{p}$ for a rigidly rotating star in the Roche approximation.} &  1 \\
$V/V_\mathrm{crit}$ & $0.8$& fixed  &5 \\ 
$V_\mathrm{crit}$ & $530\,\rm km\,s^{-1}$ & free & 1 \\
$M$ & $11.3\,M_{\sun}$  & fixed\footnote{$M$ was calculated from $V_\mathrm{crit}$.} & 1 \\
$L $ & $7400\,L_{\sun}$ & fixed & 1 \\
$T_\mathrm{eff}$ & $19\,370\,\rm K$  & fixed\footnote{$T_\mathrm{eff}$ corresponds to the average effective temperature of a rotationally deformed, gravity darkened star.} &   1 \\
\hline
\multicolumn{4}{c}{Disc Parameters}\\ 
\hline
$\Sigma_0$ & $2.1\,\rm g\,cm^{-2}$  &  free & 1 \\
$\rho_0$ & $5.9 \times 10^{-11}\,\rm g\,cm^{-3}$  &  fixed\footnote{$\rho_0$ is given by $\Sigma_0(2\pi)^{-1/2}H_0^{-1}$ (Eq.~\ref{eq:rho}).} & 1 \\
$R_\mathrm{t}$ &  $130\,R_{\sun}$  &  fixed & 1 \\
$H_0$ &  $0.208\,R_{\sun}$ & fixed\footnote{$H_0$ was calculated from Eq.~\ref{eq:scaleheight}, 
assuming  a mean molecular weight of 0.6 and an electron temperature of $18\,000\,\rm K$ after \citet{car06a}.}  & 1 \\
\hline
\multicolumn{4}{c}{$V/R$ Parameters}\\
\hline
 $P$ & 1429 d & Fixed & 2 \\
$T_0$ & 50414 \footnote{For $T_0$ we have chosen the peak value of cycle I (see Paper I).}& Fixed & 2 \\
$k_2$ & 0.006 & Free & 1\\ 
Weak line force &  $5.74\times 10^{-2}(\varpi/R_\mathrm{e})^{0.1}$ & Free & 1\\
$\dot{M}$ &$10^{-11} M_{\sun}\,\rm yr^{-1}$ & Free & 1 \\
$\alpha$  & 0.4 & Free & 1\\
$\delta$  & 0.95 & Free & 1\\
\hline
\multicolumn{4}{c}{Geometrical Parameters}\\ 
\hline
$i$ & $95\degr$ & free & 1 \\
$\gamma$ & $32\degr$ & free & 1 \\
$\phi_0$ & $280\degr$ & free & 1 \\
distance & 126 pc & fixed\footnote{$d$ was considered a free parameter within the 113 --- 148 pc range, corresponding to the accuracy in the distance determination the Hipparcos satellite.} & 6\\
\hline
\multicolumn{4}{c}{Binary Parameters}\\ 
\hline
$P_\mathrm{orb}$ & 132.97 d  & fixed & 7 \\
$e$ & 0  & fixed & 7 \\
$a$ & $257\,R_{\sun}$ & fixed  & 7\\
$q$ & 0.121 & fixed  & 7 \\
\hline
\end{tabular}
\end{minipage}
\end{table}

We assume a steady-state viscous decretion disc, for which the surface density is given by \citep[e.g.,][]{bjo05}
\begin{equation}
     \Sigma(\varpi)=\frac{\dot M}{3 \pi \alpha c_\mathrm{s}^2 }
            \left( \frac{G M}{\varpi^{3}} \right)^{1/2}
            \left[(R_0/\varpi)^{1/2}-1\right] \enspace ,
\label{eq:disc_Sigma}
\end{equation}
where
\begin{equation}
\varpi=(x^2+y^2)^{1/2}
\end{equation}
is the distance from the center of the star,
$\alpha$ is the viscosity parameter of \citet{sha73}, $\dot{M}$ is the stellar decretion rate and $R_0$ is an arbitrary integration constant. The isothermal sound speed is
\nada{$c_\mathrm{s} = (k_b T)^{1/2}(\mu m_\mathrm{H})^{-1/2}$, where $k_b$} is the Boltzman constant, $T$ is the gas kinetic temperature, $\mu$ is the mean molecular weight, and $m_H$ is the mass of the hydrogen atom.


The integration constant $R_0$ is a free parameter related to the physical size of the disc.
For time-dependent models, such as those of \citet{oka07}, $R_0$ grows with time and thus $R_0$ is related with the age  of the disc.

 $\zeta$ Tau is a well-known single line binary \citep{har84} and it is usually assumed that the tidal forces from the secondary
physically truncate the disc at a radius $R_\mathrm{t}$, corresponding to the tidal radius of the system
\citep {whi91}. We assume that the disc is sufficiently old  that $R_\mathrm{t} \ll R_0$, in which case Eq.~\ref{eq:disc_Sigma} can be written in a simple parameterized form
\begin{equation}
\Sigma(\varpi) = \Sigma_0 R_\mathrm{e}^2 \varpi^{-3/2}  \left[(R_0/\varpi)^{1/2}-1\right] 
\simeq
 \Sigma_0 \left(\frac{R_\mathrm{e}}{\varpi}\right)^2\enspace ,
\label{eq:sigma}
\end{equation}
where $\Sigma_0$ is a constant that measures the density scale at the base of the disc, and is written as
\begin{equation}
\Sigma_0 = \frac{\dot M} {3 \pi \alpha c_\mathrm{s}^2}
            \left( \frac{G M}{R_\mathrm{e}^{3}} \right)^{1/2}
            \left[(R_0/R_\mathrm{e})^{1/2}-1\right] \enspace .
\label{eq:disc_Sigma0}
\end{equation}


\nada{In Eq.~\ref{eq:disc_Sigma} the disc density scale is controlled both by $R_0$ and $\dot{M}$. 
For a truncated disc system $R_0$ is unknown, because the information about the disc age is destroyed by the tidal disruption of the outer disc by the secondary.}
Therefore, in such a system $\dot{M}$ cannot be determined observationally, unless some other information about the outflow is available. From the continuity equation
\begin{equation}
 \dot{M} = 2 \pi \varpi \Sigma(\varpi) v_\varpi
 \label{continuity}
\end{equation}
we see that the mass decretion rate is related both to the surface density and the radial velocity, $v_\varpi$. 
Thus, in order to determine the decretion rate the observations must provide a measure of the outflow velocity.

This latter quantity is difficult to obtain directly from the observations, because the outflow velocities in the inner disc are probably several orders of magnitude lower than the orbital speeds \citep{han94,han00,wat94}. 
\nada{However, the outflow velocity can, in principle,  be determined indirectly. As discussed below, the eigenvalues of the global oscillation mode depend on the viscosity parameter $\alpha$. This parameter, in turn, sets the outflow velocity (this can be seen by substituting Eq.~\ref{eq:sigma} in Eq.~\ref{continuity}).
Therefore, if we can sufficiently constrain the global oscillation eigenvalues (see Sect.~\ref{discussion}), the value of  $\alpha$ and, as a consequence, the decretion rate could be obtained.}

From the binary parameters of Table~\ref{table:model_parameters},
we find that the Roche radius of the system is $R_\mathrm{Roche}=144\,R_{\sun}$.
Here we have used the approximate formula for the Roche  
radius by \citet{egg83},
which is given by
\begin{equation}
R_\mathrm{Roche} = a \frac{0.49q^{-2/3}}{0.69q^{-2/3} + \ln (1+q^ 
{-1/3})},
\label{eq:roche}
\end{equation}
where $a$ is the semi-major axis and $q = M_{2}/M_{1}$ is the  
binary mass ratio.
Since the tidal radius is approximately $0.9\,R_\mathrm{Roche}$ in circular binaries with small mass ratio \citep[$q \la 0.3$,][]{whi91}, we assume that the disc around $\zeta$ Tau is truncated at $R_\mathrm{t} = 130\,R_{\sun}$. 

We also assume that the disc is in vertical hydrostatic equilibrium. In this case, it can be shown that the density of isothermal discs is given by \citep[][]{bjo05}
\begin{eqnarray}
  \rho(\varpi,z)&=&\rho_0(\varpi) \exp\left(-\frac{z^2}{2H^2}\right)\enspace,
\label{eq:isothermal_rho}
\end{eqnarray}
where $\rho_0(\varpi)$ is the disc density at the midplane ($z=0$), and the disc 
scale height is given by
\begin{equation}
H(\varpi) = H_0 \left( \frac{\varpi}{R_\mathrm{e}}\right)^{3/2},
\label{eq:scaleheight}
\end{equation}
where $H_0\equiv c_\mathrm{s}V_\mathrm{crit}^{-1}R_\mathrm{e}$

We can find the radial dependence of $\rho_0$ by considering that $\Sigma$ is the integral of $\rho$ in the $z$ direction
\begin{equation}
    \Sigma(\varpi)=\int_{-\infty}^\infty \rho(\varpi,z) \, dz 
          = \sqrt{2\pi}H\rho_0(\varpi) \enspace .
\label{eq:sigma_integral}
\end{equation}
From Eqs.~\ref{eq:sigma} to~\ref{eq:sigma_integral} we have
\begin{equation}
\rho(\varpi,z) = \frac{\Sigma_0}{\sqrt{2\pi}H_0}
\left(\frac{R_\mathrm{e}}{\varpi} \right)^{3.5}
 \exp\left(-\frac{z^2}{2H^2}\right)\enspace.
 \label{eq:rho}
\end{equation}

\nada{Before describing the global oscillation model}, a few words about the HDUST solution of the gas state variables are warranted. 
From Eqs.~\ref{eq:sigma} and~\ref{eq:isothermal_rho} we see that the solutions for both the vertical hydrostatic equilibrium and the radial diffusion depends  on the disc temperature. In a recent work, \citet{car08b} studied  the hydrodynamics of non-isothermal viscous Keplerian discs and found that the density structure of such discs can strongly deviate from the simple isothermal solution of Eqs.~\ref{eq:disc_Sigma} and~\ref{eq:isothermal_rho}.
The dissimilarities between the two solutions are most marked for the inner disc ($\varpi \la 10\,R_\mathrm{e}$), from whence the polarization continuum and most of the line flux comes. Indeed, a comparison between the emergent spectrum for the self-consistent solution with isothermal models revealed important differences between the two \citep[see, e.g., Fig. 6 of][]{car08b}.

The models presented in this work have an isothermal density structure but are \emph{not} isothermal, because for the assumed density distribution HDUST calculates the full NLTE and radiative equilibrium problem, thereby providing the gas temperature as a function of position in the disc.
\citet{car08b} demonstrated that such \emph{mixed models} are a much better approach to the problem than a purely isothermal model.
The reason why an isothermal density structure was assumed in this work lies in the fact that at the moment only an isothermal solution for the global disc oscillations is available. 
Lifting this inconsistency between the calculated temperature distribution and the assumed density will be left for future work.

To model the $V/R$ variations and the VLTI/AMBER observations reported in Paper I, Eq.~\ref{eq:sigma} must be modified according to the global oscillation model of \citet{oka97} and \citet{pap92}.
In calculating the gravitational potential, we take into  
account the quadrupole contribution due to the rotational deformation  
of the rapidly rotating Be primary \citep{pap92}. We also take into  
account the azimuthally averaged tidal potential \citep{hir93}.  
The potential is then given by
\begin{eqnarray}
\psi & \simeq & -\frac{GM}{\varpi} \left\{ 1 + k_{2}\left(\frac 
{\Omega_\star}{\Omega_\mathrm{crit}}
  \right)^2 \left(\frac{\varpi}{R_\mathrm{e}}\right)^{-2}
  + q \frac{\varpi}{a} \left[ 1 + \frac{1}{4}{\left(\frac{\varpi}{a} 
\right)}^2\right]
  \right\}\,,
\label{eq:potential}
\end{eqnarray}
\nada{where 
$\Omega_\star$ is the angular rotation speed of the star, $\Omega_\mathrm{crit}=2 (3R_\mathrm{p})^{-1} V_\mathrm{crit}$, and $k_{2}$ the apsidal motion constant.} 
In this equation the first term is the point-mass  
potential of the Be star,
the second term is the quadrupole contribution, and the third term is  
the azimuthally averaged tidal potential.

The radial distribution of the rotational angular velocity
$\Omega(\varpi)$ is derived
from the equation of motion in the radial direction,
and is written explicitly as
\begin{eqnarray}
     \Omega(\varpi) & \simeq & \left( \frac{GM}{\varpi^{3}} \right)^ 
{1/2}
     \left[ 1 + k_2 \left(\frac{\Omega_\star}{\Omega_\mathrm{crit}}
          \right)^2 \left(\frac{\varpi}{R_\mathrm{e}}\right)^{-2}
          - \frac{q}{2}\left( \frac{\varpi}{a} \right)^{3}
     \right. \nonumber \\
          && \left.
                 + \frac{d \ln \Sigma}{d \ln \varpi}\left(\frac{H} 
{\varpi}\right)^{2}
    -\eta \left(\frac{\varpi}{R_\mathrm{e}}\right)^\epsilon
                 \right]^{1/2},
     \label{eqn:Omega}
\end{eqnarray}
under the approximation $z^2/\varpi^{2} \ll 1$, where we have  
included a hypothetical radiative force in the form of
\begin{equation}
     F_{\rm rad} = \frac{GM}{\varpi^2}
               \eta \left( \frac{\varpi}{R_\mathrm{e}} \right)^ 
{\epsilon},
     \label{eqn:Frad_gen}
\end{equation}
where $\eta$ and $\epsilon$ are parameters characterizing
the force due to an ensemble of optically thin lines \citep{che94}.
Then the associated local epicyclic frequency $\kappa(\varpi)$ is  
written as
\begin{eqnarray}
\kappa (\varpi) & = & \left[ 2 \Omega \left( 2\Omega + \varpi \frac{d  
\Omega}{d \varpi} \right) \right]^{1/2}
\nonumber \\
       & = & \left( \frac{GM}{\varpi^{3}} \right)^{1/2}
     \left\{ 1 - k_2 \left(\frac{\Omega_\star}{\Omega_\mathrm 
{crit}}
          \right)^2 \left(\frac{\varpi}{R_\mathrm{e}}\right)^{-2}
          - 2q\left( \frac{\varpi}{a} \right)^{3}
     \right. \nonumber \\
          && \left.
          + \left[ 2\frac{d \ln \Sigma}{d \ln \varpi} + \frac{d^{2}  
\ln \Sigma}{d \ln \varpi^{2}} \right]\left(\frac{H}{\varpi}\right)^{2}
    -\eta (1+\epsilon) \left(\frac{\varpi}{R_\mathrm{e}}\right)^\epsilon
                 \right\}^{1/2}.
\label{eq:kappa}	
\end{eqnarray}

The perturbed surface density is obtained by
superposing a linear $m=1$ perturbation on the above  
unperturbed state (Eq.~\ref{eq:sigma})
in the form of normal mode of frequency $\omega$ that varies as
$\exp[i(\omega t-\phi)]$. For simplicity, the perturbation is taken  
to be isothermal.
The linearized perturbed equations are then obtained as follows,
\begin{eqnarray}
   && \left[ i(\omega-\Omega)+v_{\varpi}{d \over {d \varpi}} \right]
   {\Sigma^{\prime} \over \Sigma}
   +{1 \over {\varpi \Sigma}}{d \over {d \varpi}} \left( \varpi\Sigma v_ 
{\varpi}^{\prime} \right)
   -{{i v_\phi^{\prime}} \over \varpi}  = 0,
   \label{eqn:m1_1} \\
   && c_\mathrm{s}^2 {d \over {d \varpi}}\left( {\Sigma^{\prime}  
\over \Sigma}
   \right)
   +\left[ i(\omega-\Omega)+{{d v_{\varpi}} \over {d \varpi}}+v_ 
{\varpi}{d \over {d \varpi}}
   \right] v_{\varpi}^{\prime} - 2\Omega v_\phi^{\prime} = 0,
   \label{eqn:m1_2} \\
   && c_{\rm s}^2 \left(-{i \over \varpi}+\alpha{d \over {d \varpi}} 
\right)
   {\Sigma^{\prime} \over \Sigma}
   +{\kappa^2 \over {2 \Omega}} v_{\varpi}^{\prime} \nonumber\\
   && \hspace*{4em}
   +\left[ i(\omega-\Omega)+{v_{\varpi} \over \varpi}+v_{\varpi}{d  
\over {d \varpi}} \right]
   v_\phi^{\prime} = 0,
   \label{eqn:m1_3}
\end{eqnarray}
where 
$\Sigma^ 
{\prime}$ is
the Eulerian surface-density perturbation, and $(v_{\varpi}^{\prime},  
v_\phi^{\prime})$
is the vertically averaged velocity field associated with the
perturbation.
We solve Eqs.~(\ref{eqn:m1_1})-(\ref{eqn:m1_3}) with a rigid wall  
boundary condition,
$(v_{\varpi}^{\prime}, v_\phi^{\prime})=0$, at the inner  
edge of the disc
and a free boundary condition, $\Delta p=0$, at the outer disc
radius, where $\Delta p$ is the Lagrangian perturbation of pressure.

It is instructive to consider the effects of some parameters on the $m=1$
oscillations before solving the perturbation equations. For
simplicity, we first neglect viscous terms. Then, the local dispersion
relation is obtained from Eqs.~(\ref{eqn:m1_1})--(\ref{eqn:m1_3}) as
\begin{equation}
   (\omega - \Omega - k v_{\varpi} )^2-\kappa^2 = c_\mathrm{s}^2 k^2,
   \label{eqn:confine}
\end{equation}
where $k$ is the radial wave number. In this case, the propagation
region, where $k$ is a real number, is given as the region where 
$(\omega-\Omega)^2-\kappa^2+v_{\varpi}^2\kappa^2/c_\mathrm{s}^2>0$
\citep[e.g.,][]{oka00}, which is explicitly written as
\begin{eqnarray}
   \omega &<& \Omega - \kappa \left[1-\left(\frac{v_{\varpi}}{c_\mathrm{s}}
       \right)^2 \right]^{1/2} \nonumber \\
   &\simeq& \left(\frac{GM}{\varpi^{3}}\right)^{1/2}
        \left[ -\frac{1}{2}\left(\frac{d \ln \Sigma}{d \ln \varpi}
         +\frac{d^{2} \ln \Sigma}{d \ln \varpi^{2}}\right) 
         \left(\frac{H}{\varpi}\right)^{2}
         +\frac{3}{4}q \left(\frac{\varpi}{a}\right)^{3} 
         \right. \nonumber \\
   && \left. +k_2 \left(\frac{\Omega_\star}{\Omega_\mathrm{crit}}\right)^
2
         \left(\frac{R_\mathrm{e}}{\varpi}\right)^{2}
         +\frac{1}{2}\eta \epsilon
	 \left(\frac{\varpi}{R_\mathrm{e}}\right)^\epsilon
         +\frac{1}{2}\left(\frac{v_{\varpi}}{c_\mathrm{s}}\right)^2 \right].
\label{eqn:prop-region}
\end{eqnarray}
Here, the first, second, and third terms are due to the pressure
gradient force, tidal force and the quadrupole contribution of the
potential around the deformed star, respectively. The
fourth term is due to the radiation force and the last term is the
contribution \nada{due to advection}. Near the star, the quadrupole term and
the radiation term play the major role. Thus, the
eigenfrequency is mainly determined by these terms and the pressure
gradient term. The larger the contribution of the quadrupole term and
the radiation term and/or the smaller the sound speed (or
temperature), the higher the eigenfrequency. On the other hand, the
other terms, i.e., the tidal term and the advective term, as well as
the pressure gradient term affect the oscillation behavior in the
outer part of the disc. The larger the contribution of these terms,
the less the eigenmode is confined, and vice versa.

Next, we consider the effects of viscosity. \nada{When} the
mass-decretion rate $\dot{M}$ and the surface density at the inner
disc radius $\Sigma_0$ are fixed, the radial velocity $v_{\varpi}$ is
proportional to the viscosity parameter $\alpha$. Thus, one effect of
viscosity is to make the oscillation mode less confined, in terms of
the advective term. 
\citet{kat78} showed that $m=1$ modes in nearly
Keplerian discs, which are neutral if the viscosity is neglected,
become overstable when \nada{viscous effects} are taken into account. The
growth rate is proportional to $\alpha$ \citep[see also][]{neg01}.

It is important to note that our model provides prograde modes for a
plausible range of parameters. The fundamental mode, in general, has
the eigenfrequency of the order
$\Omega(R_\mathrm{e})-\kappa(R_\mathrm{e})$. From
Eq.~(\ref{eqn:prop-region}) it is obvious  that
$\Omega(R_\mathrm{e})-\kappa(R_\mathrm{e})$ is positive even if no
radiation force is included. \nada{Thus, there is} no way to reproduce the
observed V/R period of $\zeta$ Tau by retrograde modes, which have
negative eigenfrequencies.

It should also be noted that linear models like ours \nada{cannot predict the amplitude of the oscillations}.
Thus, for the purpose of comparison with various observations,
we will assume that the nonlinear perturbation patterns are similar to
the linear eigenfunctions obtained from the above equations,
and take the maximum value of the perturbed part of the surface density,
$\Sigma^{\prime}/\Sigma$, to be a free parameter, $\delta$.

\section{Results\label{results}}

\subsection{Modeling Procedure}

\nada{The global oscillation model described above is not a trivial problem for a radiative transfer solver. 
A very large number of grid cells (typically about $100\,000$) is needed to accurately describe the $\varpi$, $z$ and $\phi$ dependence of the density, of the gas state variables and of the radiation field \citep[for details about the adopted cell structure in HDUST, such as radial and latitude spacing, see][]{car06a}.
Solving the radiative equilibrium and NLTE statistical equilibrium problems for each cell involves a long series of iterations to accurately determine the coupling between radiation and state variables in the dense environment of the inner disc. This, and the subsequent calculation of the observables, takes between 2 and 3 days for a single model in a cluster of 50 Pentium IV computers.}

This is prohibitively long for the iterative procedure necessary to fit the observations. 
For this reason, we have initially modeled the global disc properties using a simple 2D viscous decretion model using Eqs.~(\ref{eq:sigma}) and~(\ref{eq:isothermal_rho}). 
This 2D analysis allowed us to place initial constraints on several parameters of the system, including $\Sigma_0$, $V_\mathrm{crit}$, $i$, and $d$ (\nada{Sect.~\ref{2dmodel})}.

Once a suitable 2D model was found, we used it as a starting point for the detailed 3D modeling described in Sect.~\ref{disc_structure}. 
With this 3D model \nada{we performed a simultaneous analysis of the VLTI/AMBER interferometry and the $V/R$ properties of the \ion{H}{I} lines, that allowed us to constrain well the global disc oscillations parameters (Sect.~\ref{3dmodel}).}

\subsection{Two-Dimensional Viscous Decretion Disc Model \label{2dmodel}}

   \begin{figure}
   \centering
   \includegraphics[width=\columnwidth]{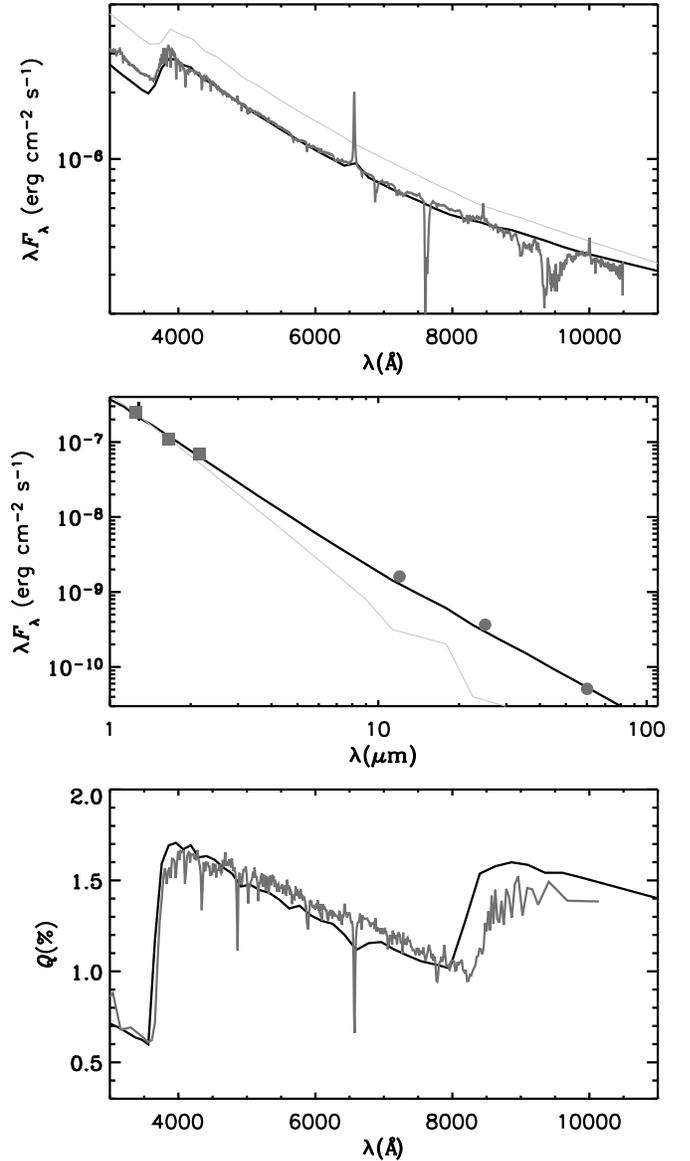}
      \caption{Emergent spectrum of $\zeta$ Tau. The dark grey lines and symbols are the observations and \nada{the black lines represent the 2D model results}.
    \emph{Top: } Visible SED \citep[data from][scaled in flux to match the average $V$-band magnitude from 1992 to the present]{woo97}.
   \emph{Middle:} IR SED (data from 2MASS -- squares, and IRAS -- circles).
   \emph{Bottom: } Continuum polarization \citep[data from][also scaled in level to match the average $V$-band polarization from 1992 to the present]{woo97}.
    In the two upper panels, the light grey line corresponds to the unattenuated stellar SED
        }
    \label{mosaic2d}
   \end{figure}
 %

   \begin{figure}
   \centering
   \includegraphics[width=\columnwidth]{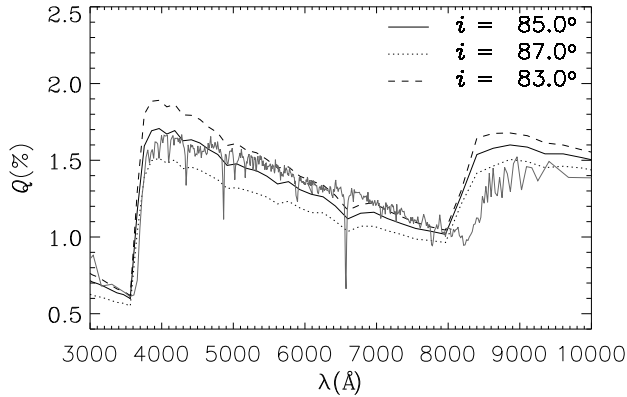}
      \caption{
      Continuum polarization for $\zeta$ Tau.  The model results for three different viewing angles, as indicated, are compared with the observations
              }
    \label{i_check}
   \end{figure}

\nada{As discussed in Paper I, $\zeta$ Tau has} shown little or no secular evolution since the onset \nada{of} the current $V/R$ activity \nada{started} in 1992, and it is argued that the global disc properties, basically its density scale, have been approximately constant in the period.
Therefore, modeling the data with a 2D viscous decretion disc is per se \nada{a useful} exercise that can provide insight into the average properties of \nada{this disc}.

\nada{Initially, a set of observations that are representative of the period from 1992 through 2008 must be defined.}
For both the spectral energy distribution (SED) and polarization in the $0.3$ --- $1\,\mu\rm m$ region, we used the PBO spectropolarimetric observations made in March, 1996 \citep{woo97}. The flux and polarization levels were scaled in order to match the average $V$-band flux and polarization of the period ($\langle V\rangle=3.02 \pm 0.07$, $\langle p_V\rangle=1.46\pm0.08$\%, see Paper I). To extend the SED to the IR, we used archival data from the IRAS point source catalog \citep{bei88} and the 2MASS catalog \citep{2mass}.

Because the line profiles are variable and strongly dependent on the disc asymmetries caused by the global oscillations, we have fitted only the average H$\alpha$ peak separation in this period ($240\rm km\,s^{1}$, Paper I) 
and the average H$\alpha$ EW ($-15.5\pm0.1\,\rm\AA$, Paper I).

One of the most beautiful characteristics of the viscous decretion disc model is its relatively small number of parameters. To describe the steady-state structure of the disc we need to specify the decretion rate, $\dot{M}$, the viscosity parameter, $\alpha$, and the age of the disc that is enclosed in the parameter $R_0$ (Eq.~\ref{eq:disc_Sigma}). We further need to specify the stellar critical velocity, $V_\mathrm{crit}$, which is a measure of the stellar mass and affects the disc structure by setting its scaleheight (Eq.~\ref{eq:scaleheight}) and rotation speeds.
In the case of the disc around $\zeta$ Tau we assume,  as explained in Sect.~\ref{disc_structure}, that the disc is truncated by \nada{a companion star. This mechanism erases} the information about the disc age and makes the problem undetermined. This basically means that $\dot{M}$, $\alpha$ and $R_0$ are all contained in a single parameter, $\Sigma_0$, which determines the disc density scale.
Therefore the \nada{disc} is described by only two free parameters, $\Sigma_0$ and $V_\mathrm{crit}$. 
In addition to \nada{these} two physical parameters we must specify two geometrical parameters, the viewing angle, $i$, and the distance to the star, $d$.

For this analysis several tens of models were run, covering a large range of values for $\Sigma_0$, $V_\mathrm{crit}$ and $i$. 
For each model, the value of $d$ was computed in order to match the observed flux levels.
Since $d$ is known to be in the range between $113$ --- $148\rm\,pc$, which corresponds to the accuracy in the distance determination by the Hipparcos satellite \citep{per97}, models that did not produce $d$ in this range were discarded. 

Figs.~\ref{mosaic2d} shows our best fit to the data using the 2D viscous decretion model. This best fit was achieved for the
following parameters
\begin{list}{}{}
\item[] $\Sigma_0 = 1.7\rm\,g\,cm^{-2}$\,,
\item[] $i = 85\degr$\,,
\item[] $V_\mathrm{crit} = 530\rm\,km\,s^{-1}$\,,
\item[] $d = 126\,\rm pc$\,,
\end{list}
in addition to the fixed stellar parameters listed in the first part of Table~\ref{table:model_parameters}. We recall that for all models the disc was truncated at $130\,R_{\sun}$.

The above value for $\Sigma_0$ corresponds to $\rho_0 = 5.9\times 10^{-11}\,\rm g\,cm^{-3}$. \nada{The number density of particles can be estimated assuming a mean molecular weight of 0.6, typical for ionized gas with solar chemical composition (recall that our models have only H).}
We obtain a value of $5.9 \times 10^{13}\rm\,cm^{-3}$, which is characteristic of Be stars with dense discs, such as $\delta$ Sco \citep[][]{car06b}.

Before discussing our results, it is useful to recall the physical processes that control the emergent spectrum.
In the visible, the SED depends on the spectral shape of the stellar radiation, which, for a rotating star, is controlled by the latitude-dependent effective temperature.
The stellar radiation is modified by several processes in the CS disc: electron scattering, bound-free absorption and free-bound emission by \ion{H}{I} atoms, free-free absorption and emission by free electrons, \nada{and} line absorption and emission by H and other elements.
The absorption and emission processes depend on the atomic occupation numbers and electron temperature, which, in turn, depend on the radiation field in a non-linear and complex way.

In the IR the SED is controlled mainly by the free-free emission of the \nada{free} electrons.
The continuum polarization is produced by electron scattering, but is also modified  by the absorption and emission processes in the disc. 

The absorption, emission and scattering of the radiation depends also on the geometry and velocity structure of the CS material. For instance, the linear polarization is strongly dependent \nada{on the number density of electrons, the disc flaring and scaleheight. The slope of the IR SED, on the other hand, depends mainly on the radial profile of the density, but also on the disc flaring and temperature distribution \citep{car06a}.}

In view of the complex dependence of the emergent radiation on the stellar and disc parameters, the overall good match between the model and observations is remarkable. The model reproduces the SED longward of the Balmer jump ($3646\,\rm\AA$) all the way down to the longest IRAS wavelength. 
The only part of the SED which is not well reproduced is the size of the Balmer jump. In our models, the size of the jump is too large, meaning that either the amount of neutral hydrogen in the line of sight is a little overestimated or the geometry of inner disc is not well described in our models (see Sect.~\ref{discussion}). 

\nada{The predicted polarization matches the observations well. The size of the Balmer jump and the level and slope of the polarization in the Paschen continuum show good agreement. 
However, in the Brackett continuum our models overestimate the polarization by approximately $10\%$.}

\nada{We obtain an inclination angle of  $i=85\degr$ for our best-fit model. 
The fact that  $\zeta$ Tau is a well-known shell star \citep{por96,riv06} supports our findings. 
In Fig.~\ref{i_check} we compare the model polarization for 3 different inclination angles, 83\degr, 85\degr and 87\degr. Differences in $i$ as small as 2\degr~produce significant changes in the polarization,  which indicates that the inclination angle is well constrained by our analysis.}

The extent \nada{that} the CS disc modifies the stellar radiation can be assessed by comparing the model SED with the unattenuated stellar radiation in Fig.~\ref{mosaic2d}. 
Since the star is viewed close to edge on, this causes extinction of UV and visible radiation \nada{that is reemitted at} longer wavelengths. Most of the reprocessed radiation escapes in the polar direction, due to the fact that the vertical optical depth is much smaller than the radial optical depth. \nada{However, a portion of this radiation will escape radially and will contribute to the large IR excess observed} ($\approx 3\,\rm mag$ at $25\,\rm\mu m$). 

The H$\alpha$ peak separation and EW for the best-fit model are  $240\rm km\,s^{-1}$ and $-14\,\rm\AA$, respectively, in good agreement with the average values \nada{for this time period}.
We have chosen $V_\mathrm{crit}$ as a free parameter and, \nada{therefore}, no a priori assumption about the stellar mass have been made.
$V_\mathrm{crit}$ is well constrained by the model, \nada{as both the CS emission lines and the continuum polarization depend on this value (recall that $V_\mathrm{crit}$ affects the disc scaleheight)}.
\nada{$V_\mathrm{crit}$ and the assumed $R_\mathrm{e}=7.7\,R_{\sun}$ give a stellar mass of $11.3\,M_{\sun}$, in agreement with \citet{har84}. This further supports the values of the parameters adopted for $\zeta$ Tau}.

\nada{The good fit that was obtained using a relatively simple model provides strong observational support to our assumptions.}
The power-law description for the surface density proved successful in determining the slope of the IR SED, and the hydrostatic equilibrium assumption successfully determined the correct shape and level of the continuum polarization.
\nada{Therefore, we conclude} that a steady-state viscous decretion disc, truncated at $R_{\rm t} = 130\,R_{\sun}$, is a good description for the average properties of the CS disc around $\zeta$ Tau. 

\subsection{Global Oscillation Model \label{3dmodel}}

   \begin{figure}
   \centering
   \includegraphics[width=\columnwidth]{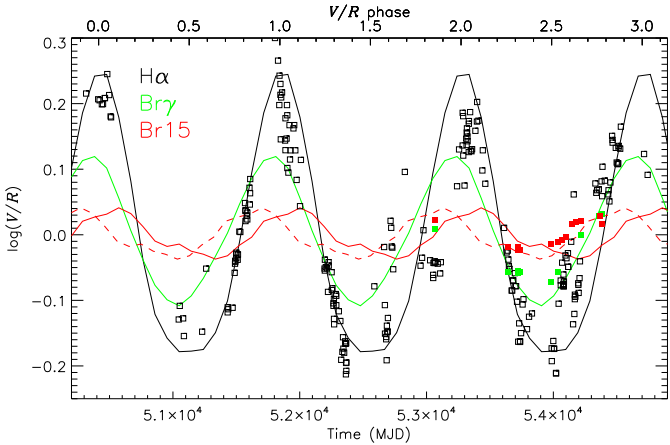}
      \caption{
      $\log(V/R)$ vs. time for the H$\alpha$ (black), Br$\gamma$ (green) and Br15 (red) lines. The points correspond to the observations, and the lines to our best fitting model.
      The red dashed line corresponds to the Br15 $V/R$ curve \nada{artificially shifted in phase by -0.2 to match the observations}
              }
         \label{vrphase}
   \end{figure}

\nada{We now present our analysis of the VLTI/AMBER observations and the $V/R$ properties of $\zeta$ Tau using the global oscillation model. 
In addition to the parameters $\Sigma_0$, $V_\mathrm{crit}$, $d$, and $i$, that we have discussed above, we now consider the parameters $k_2$, $\delta$, $\dot{M}$, $\alpha$ and the weak line force, so that we can consider the $m=1$ density perturbation within the disc.
Also, the time-dependent position of the $m=1$ perturbation pattern with respect to the direction of the observer, described by the parameter $\phi$, must be determined (see Fig.~\ref{geometry}). }



Different aspects of our solution are illustrated in Figs.~\ref{vrphase} to \ref{pphase}, and the 
parameters of the best fit obtained are summarized in Table~\ref{table:model_parameters}.
\nada{Before continuing to describe our results, some explanations about our modeling procedure are warranted.}
For a given model, HDUST computes the emergent spectrum for an observer whose position is specified by $i$ and $\phi$.
To simulate the temporal dependence of the observables as the density wave precesses around the star, we compute, for a given $i$, the emergent spectrum for several values of $\phi$. 
The global oscillation parameters used in the simulation are then evaluated by comparing the theoretical amplitude and shape of the $V/R$ cycle to the observations.

The time-dependent position of the density wave is related to the $V/R$ phase as follows.
According to the convention of Paper I, we define the $V/R$ phase $\tau$ such that $\tau=0$ at the $V/R$ maximum. The ephemerides of the $V/R$ cycle, given by the period, $P$, and the modified Julian date for $\tau= 0$, $T_0$, were determined in Paper I and are given in Table~\ref{table:model_parameters}.
We introduce the parameter $\phi_0$, which gives the position of an arbitrary point of the spiral pattern at $\tau = 0$.  We chose this point to be the minimum of the density perturbation pattern at the base of the disc.
The angle $\phi$ is related to the phase by 
\begin{equation}
\tau = 1-\frac{\phi-\phi_0}{2\pi}.
\end{equation}
(Note that the angle $\phi$ \emph{decreases}  with time for the prograde oscillation mode assumed here).
Since the phase is related to the time by $\tau = (T-T_0) P^{-1}$, we find the following expression for the temporal dependence of $\phi$
\begin{equation}
\phi = 2\pi\left(1-\frac{\mathrm{T}-\mathrm{T}_0}{P}\right) + \phi_0.
\label{phi_T}
\end{equation}
\nada{This equation} gives the position of the minimum of the density pattern at the base of the star as a function of time. For a given model, a value of $\phi_0$ is obtained by fitting the $V/R$ phase and the interferometric observations.

\subsubsection{Constraints from the $V/R$ Observations}

In Fig.~\ref{vrphase} we show our best fit to the $V/R$ cycle of the H$\alpha$, Br$\gamma$ and Br15 lines. 
For H$\alpha$, the model reproduces the first part of the $V/R$ cycle ($0<\tau<0.5$), including the shape of the curve and the observed $V/R$ maxima and minima.
In the second half of the cycle, between $0.5<\tau<0.8$, the observed $V/R$ ratio departs from the quasi-sinusoidal shape that is characteristic of the first half (Paper I), and this is not reproduced by the model.

The behavior of the $V/R$ properties of the IR lines are similar to \nada{that} of H$\alpha$, but the smaller number of observations make the results less conclusive.
Unfortunately,  there are only three observations in the first half of the cycle, but their $V/R$ ratios agree with the predicted values (see points for $\rm MJD\approx 53\,700$).
The second half of the cycle is \nada{better} covered. 
For Br$\gamma$, the model seems to reproduce the data, but the $V/R$ values tend to cluster above the predicted curve, \nada{with the exception of} the two points for $\tau > 0.8$.
For Br15, all the data points are above the predicted values in the second half of the cycle.

\begin{figure}
   \centering
\includegraphics[height=3.8cm]{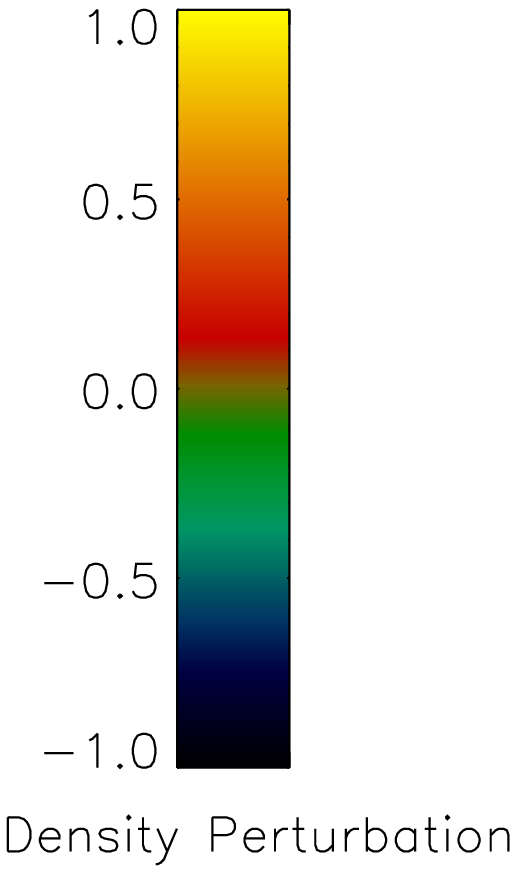}%
\includegraphics[width=3.8cm]{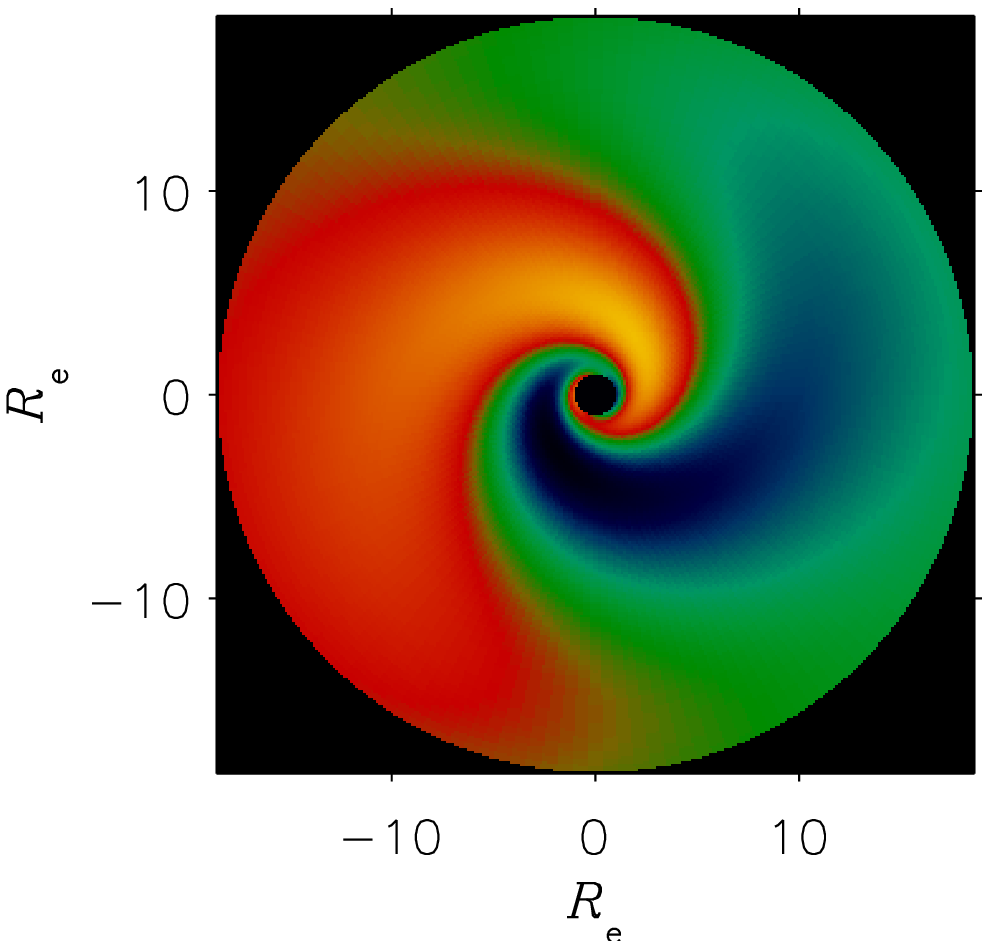}%
\includegraphics[width=3.8cm]{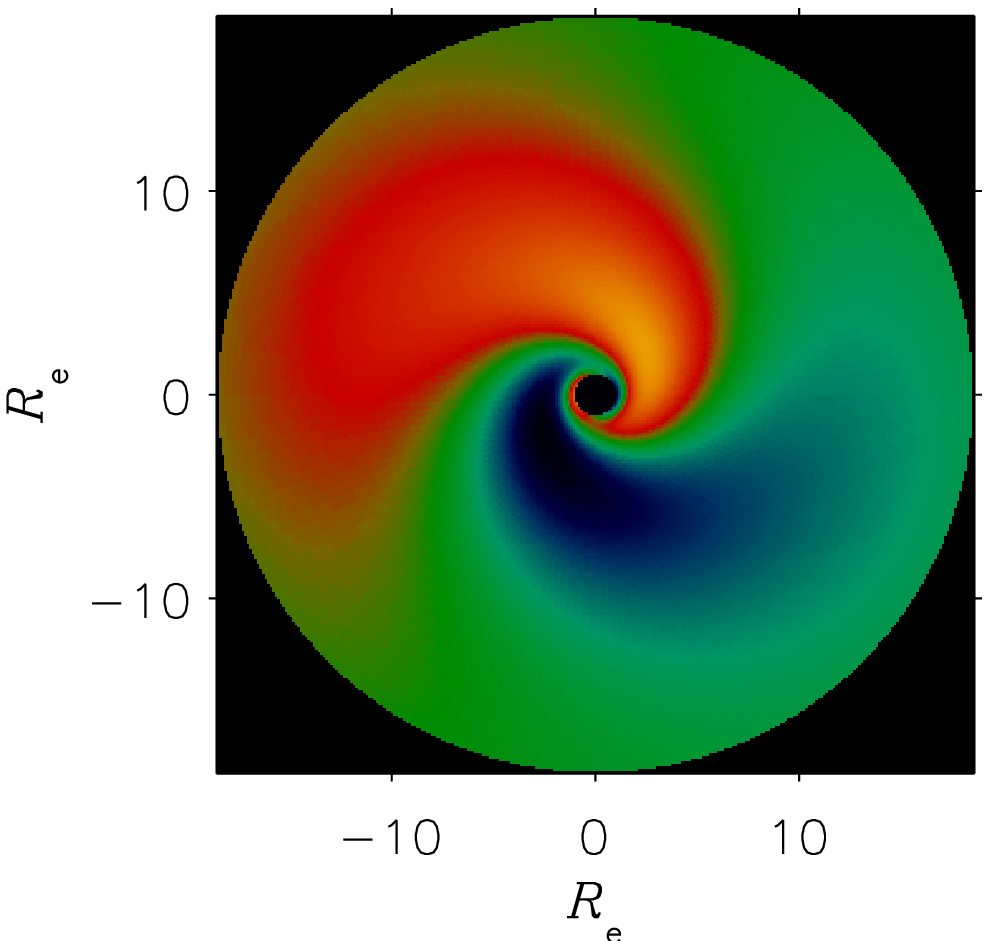}%

\medskip					  
\includegraphics[width=8.0cm]{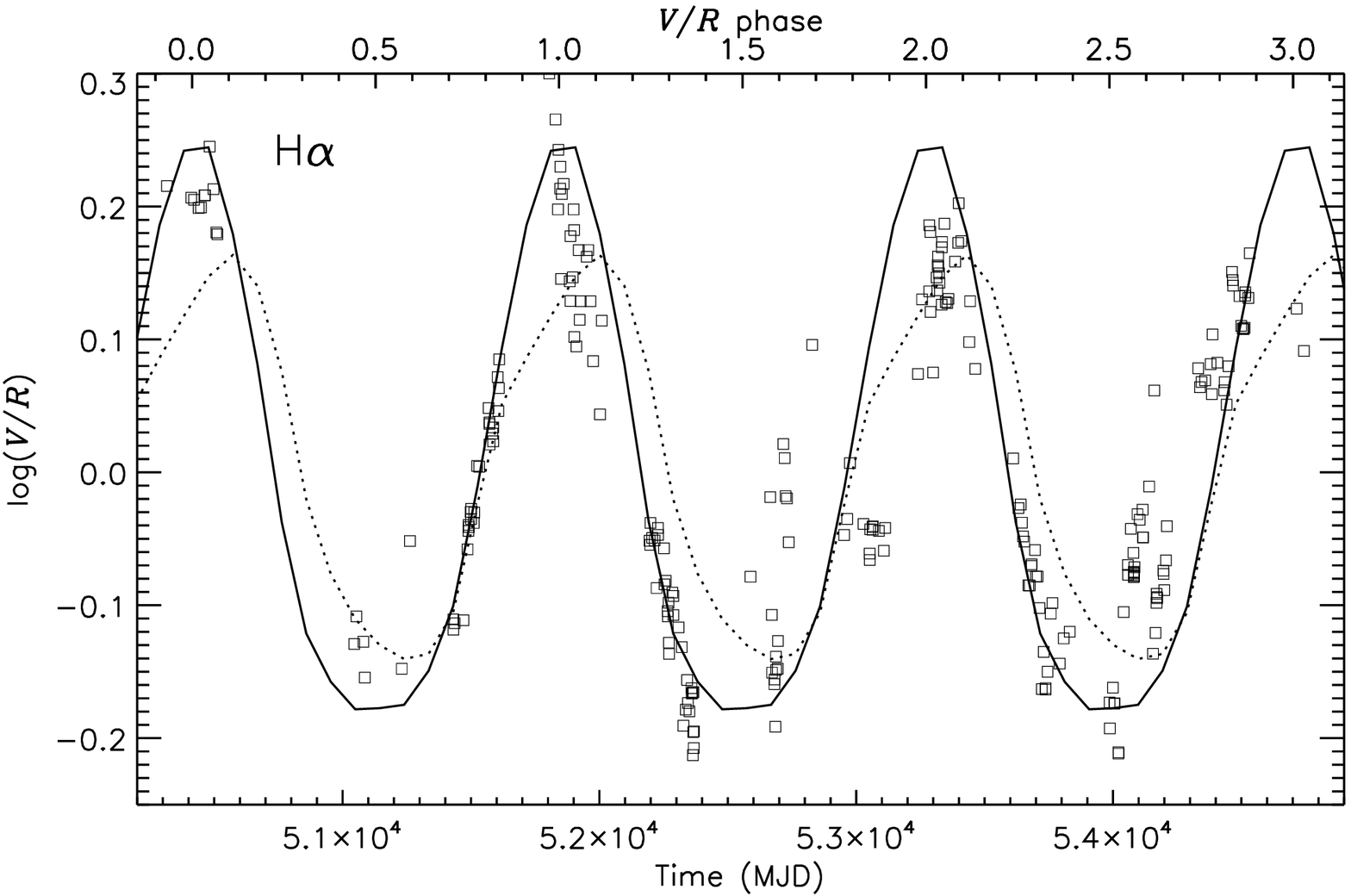}
      \caption{Comparison of the results for modes with different degrees of confinement.
      The top left plot shows the density perturbation pattern for the $V/R$ parameters of Table~\ref{table:model_parameters}. 
      \nada{The picture depicts the disc as seen from above and direction of precession of the density wave is counter-clockwise.}
      A \nada{more} confined mode, with a weak line force of  $0.04(\varpi/R_\mathrm{eps})^{0.1}$, $\alpha=0.3$ and $k_2= 0.006$, is depicted in the top \nada{right} panel.
      The bottom panel shows the predicted amplitude of $V/R$ ratio for each model (solid line: less confined model; dotted line: more confined model).
             }
         \label{confinement}
   \end{figure}

One interesting feature of the observations \nada{that is} clearly seen in Fig.~\ref{vrphase}, 
are the phase differences between the cycles of different lines. These phase differences have been already observed in $\zeta$ Tau by \citet[]{wis07}, who suggested
that they might be associated with the helicity of the global oscillation. 
The idea behind this suggestion is that \nada{the more optically thin infrared lines form closer to the star in the disc, whereas the H$\alpha$ line forming region extends further out into the disc}. Because of the strong helicity of the density perturbation pattern (see, e.g., Fig.~\ref{confinement}), the average azimuthal morphology of the inner disc is quite different than the morphology of the outer disc.
\nada{Therefore, the observed phase differences would be a result} of the different average morphology of each line formation site.

The phase differences between the H$\alpha$ and Br$\gamma$ lines are well reproduced by our models, which provides convincing evidence in support to a spiral morphology of the density perturbation pattern in the disc of $\zeta$ Tau.
The phase of the Br15 line is not well reproduced, however. 
This line is formed in the very inner part of the disc and it seems, therefore, that the global oscillation model does not describe correctly the shape of the spiral pattern in this region.
In Fig.~\ref{vrphase}, the dashed red line shows the model $V/R$ curve of the Br15 line shifted in phase by -0.2. This artificially displaced curve \nada{reproduces} the observations.
This may indicate that the model \nada{predicts} the correct amplitude of the oscillations in the inner disc, even though the shape, which sets the phase, is not correct.

As mentioned above, the global oscillation model failed to reproduce the H$\alpha$ $V/R$ cycle for 
$0.5\lesssim \tau \lesssim0.8$. 
\nada{However, this} may not be a problem with the global oscillation model, since the 
$0.5\lesssim \tau \lesssim0.8$
 phase interval is where the complex \nada{triple-peaked H$\alpha$ profiles become more prominent.
In Paper I, we have presented evidence that the triple-peak anomaly arises from the outermost part of the disc, and that it is probably a line-of-sight effect owing to the large inclination angle of the system.}
The failure to reproduce this part of the $V/R$ cycle is \nada{likely} due to the fact that a physical process is missing in our analysis.

Fig.~\ref{confinement} illustrates the properties of our solution for the disc global oscillation.
The oscillation mode is subject to two very stringent constraints: it must have a precessing period equal to the observed $V/R$ period (1\,429 days, see Paper I) and it must be able to reproduce the observed amplitude of the $V/R$ variations.
This latter requirement is associated with the \emph{degree of confinement of the mode}, which essentially means how far out into the disc the perturbation goes.
In Fig.~\ref{confinement} we compare the theoretical $V/R$ curve for two modes with the same precession period but different degrees of confinement.
The model shown in the top left is the best fit model of Table~\ref{table:model_parameters}. This model corresponds to \nada{a mode with a small degree of confinement}, in which the oscillations extend all the way down to the outer rim of the disc. The other model is shown as an example of a mode that also has the same precession period but it is much more confined. 
From the comparison of the predicted $V/R$ ratios for each mode (bottom panel of Fig.~\ref{confinement}) we see that this latter mode can be readily discarded on the basis that it cannot account for the observed amplitude of the $V/R$ variations.

 %


\subsubsection{Contraints from the AMBER/VLTI Observations}

\begin{figure*}[htb]
   \centering
    \includegraphics[width=13cm,clip,angle=270]{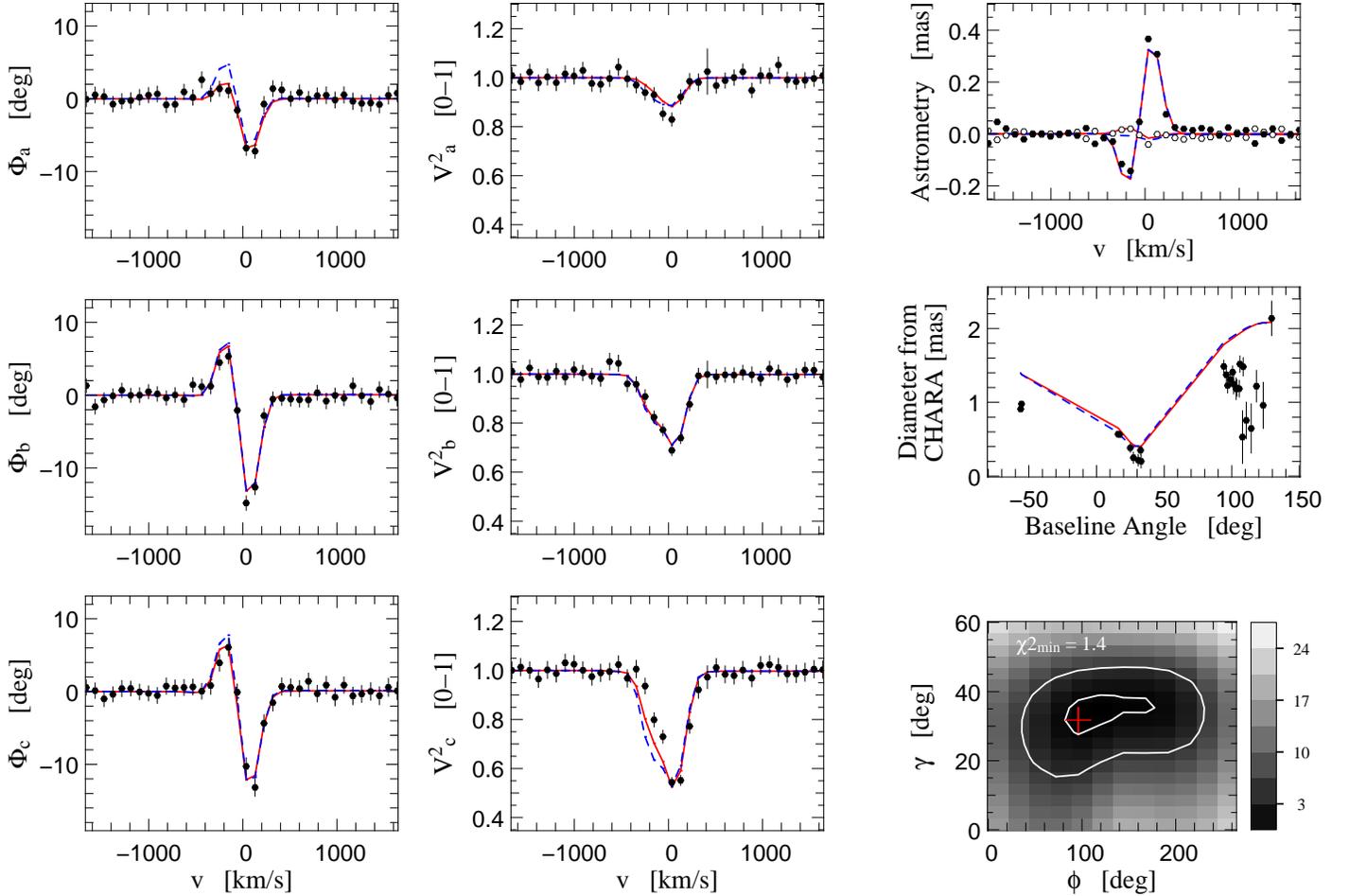}
       \caption{
       Fitting of the interferometric observations \nada{for the spectral region around the Br$\gamma$ line}. In all panels the solid line represents our best-fit model of Table~\ref{table:model_parameters} and the dashed line a model with $i=85\degr$.
        {\it Left and Middle:} Differential phases and differential visibilities obtained with AMBER/VLTI, for three interferometric baselines. The baseline orientations are indicated in Fig.~\ref{images}.
         {\it Top-Right:} Interferometric phases converted into astrometric shift (photocenter displacement) vs. wavelength. Open and filled symbols represent displacement along the disc minimum and maximum elongation axis respectively.
       {\it Middle-Right:} Angular diameter derived from a Gaussian fit of the CHARA $K^{\prime}$ band continuum observations of \citet{gie07}.
       {\it Bottom-Right:} $\chi^2$ map showing the fit for the $\gamma$ and $\phi$ angles \nada{for the  $i=95\degr$ model, with contours plotted for $\chi^2=4$ and $\chi^2=6$}
       }
         \label{visib}
   \end{figure*}

\begin{figure*}
   \centering
   \includegraphics[height=5.2cm]{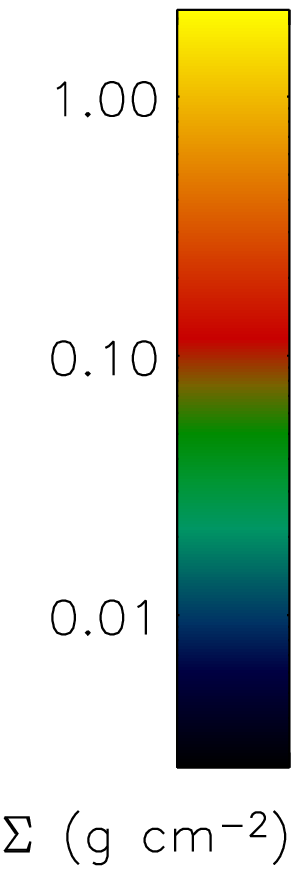}
   \includegraphics[width=5.2cm]{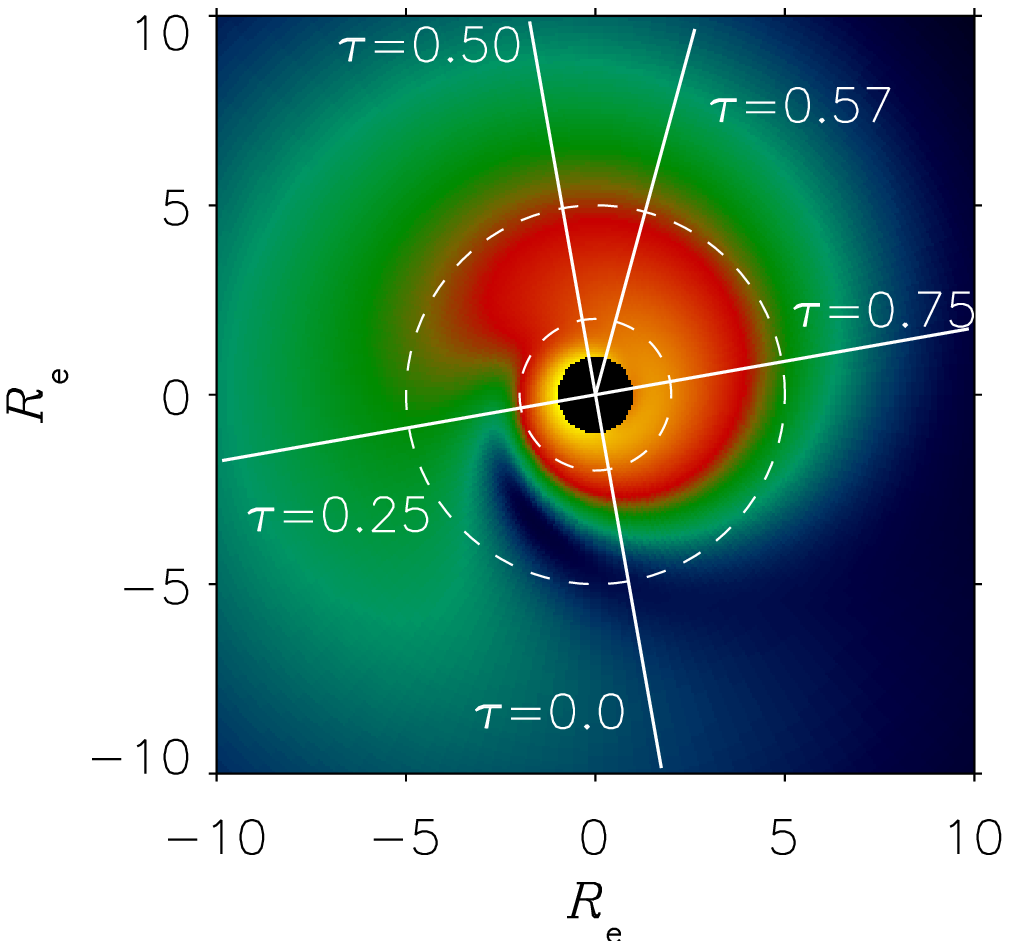}
   \includegraphics[width=5.2cm]{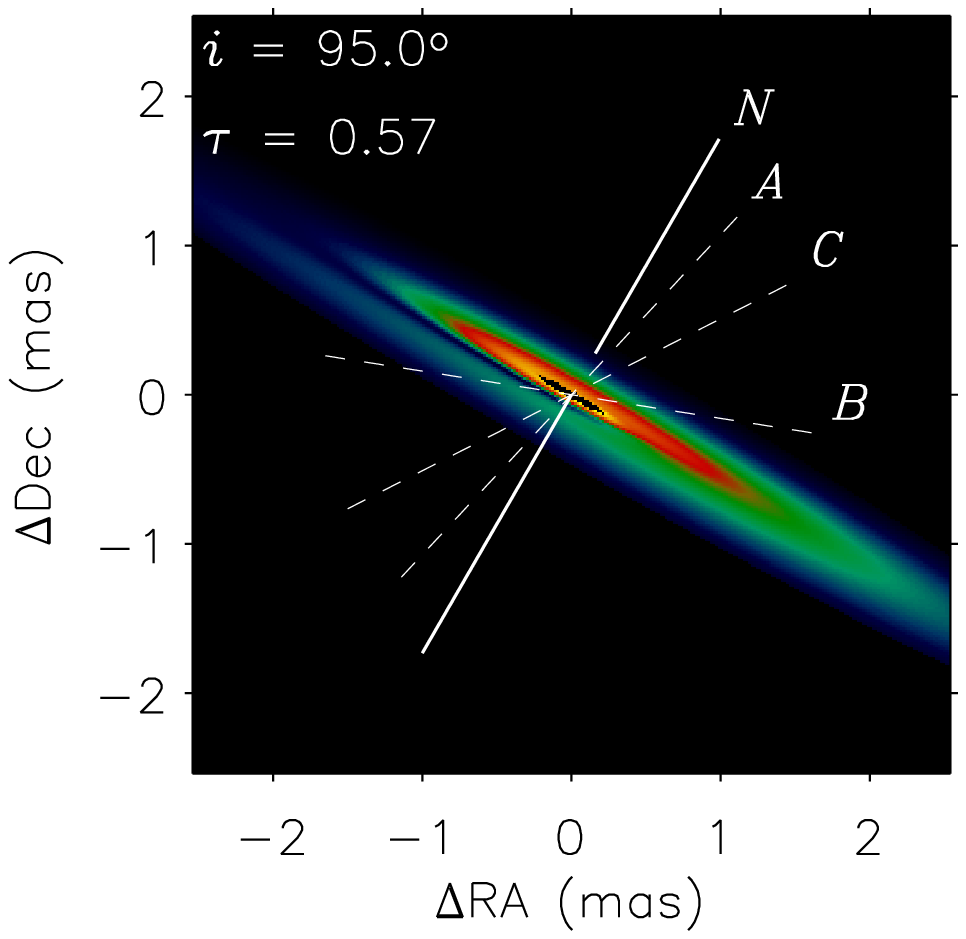}
   \includegraphics[width=5.2cm]{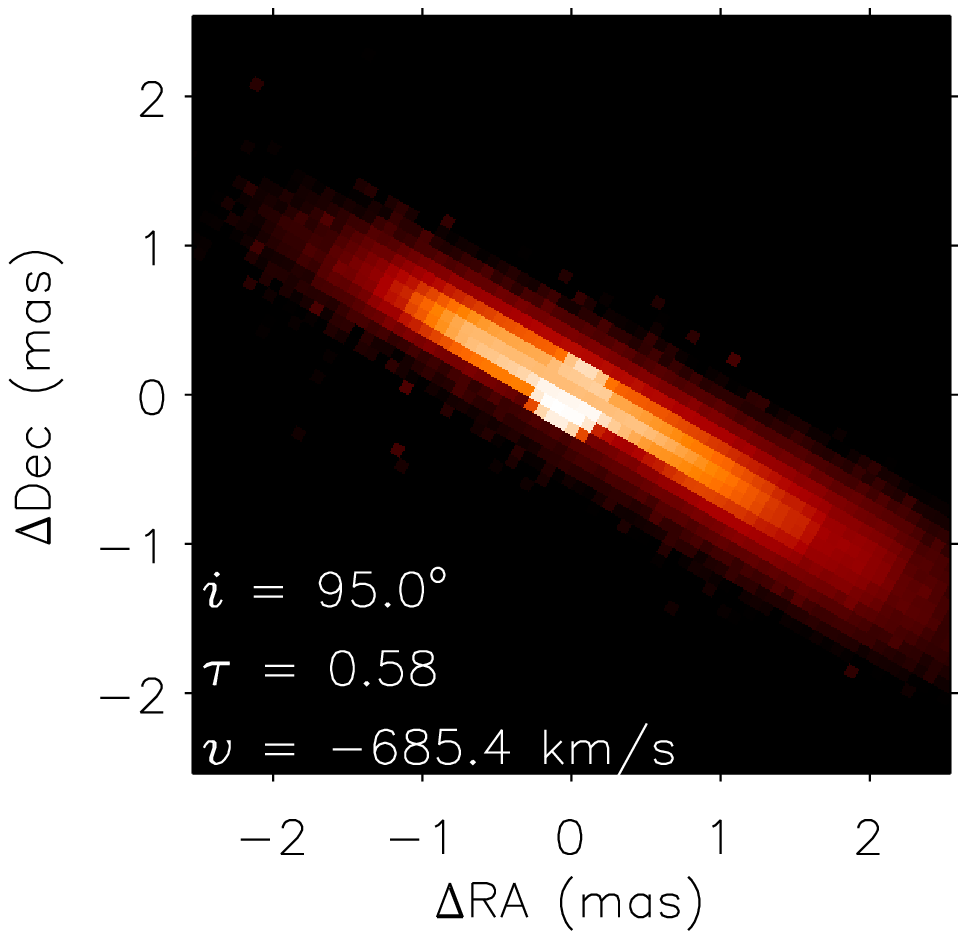}
   \includegraphics[height=5.2cm]{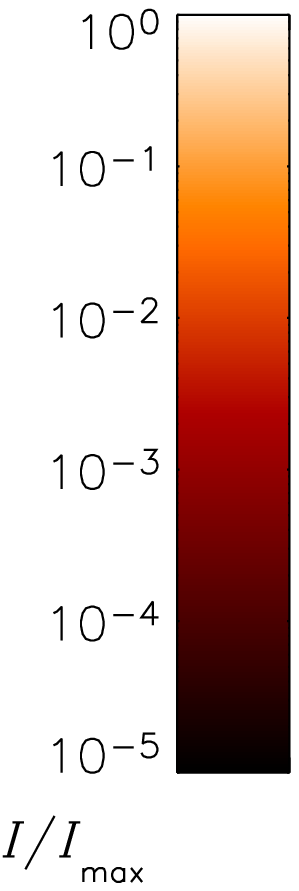}
   \includegraphics[width=5.2cm]{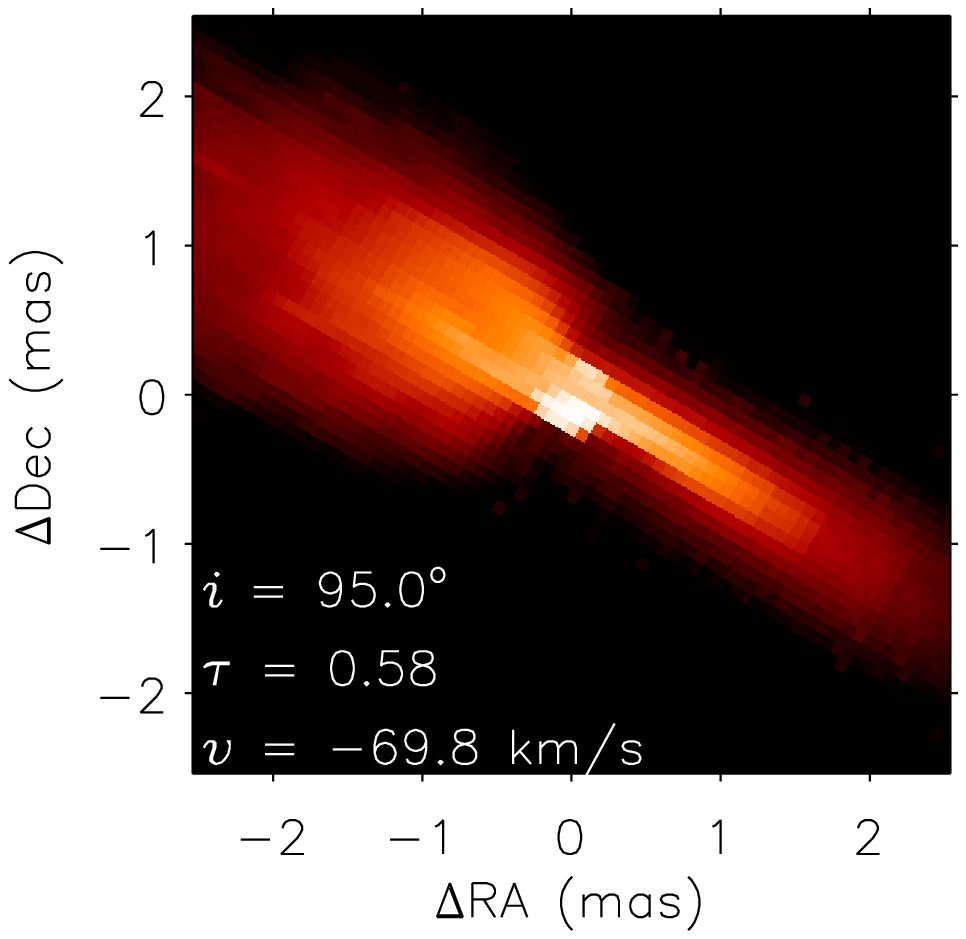}
   \includegraphics[width=5.2cm]{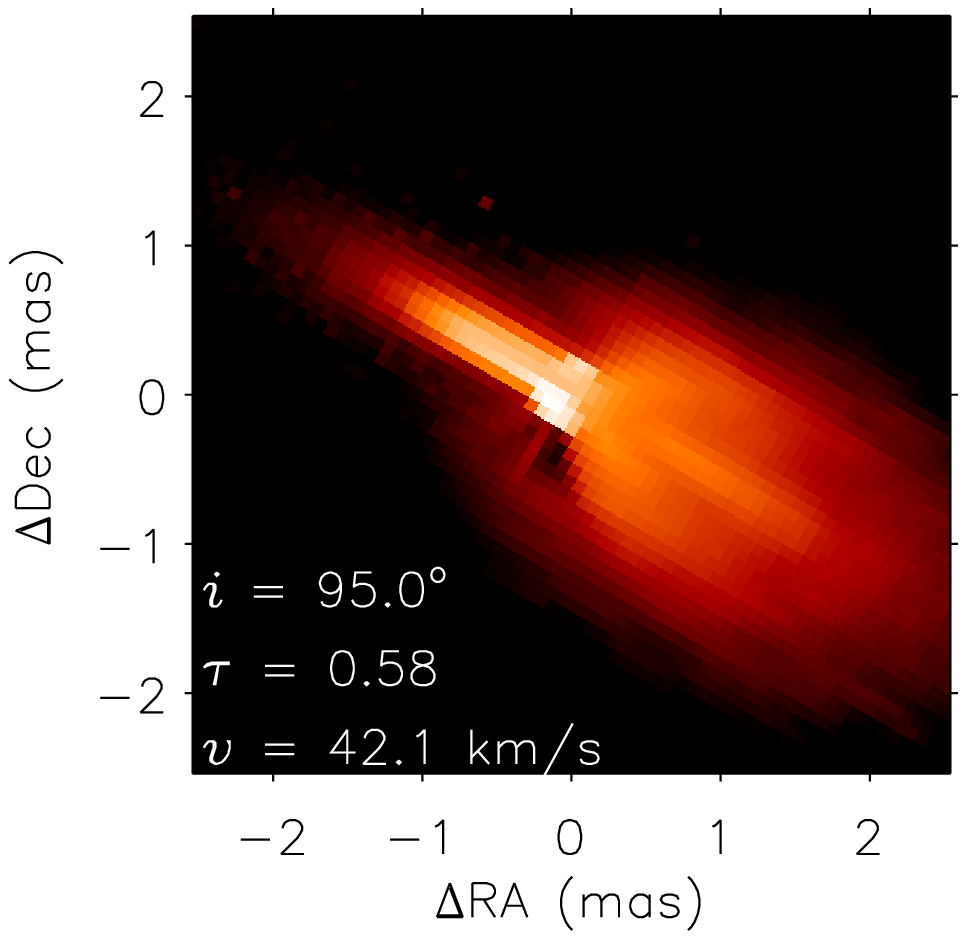}
   \includegraphics[width=5.2cm]{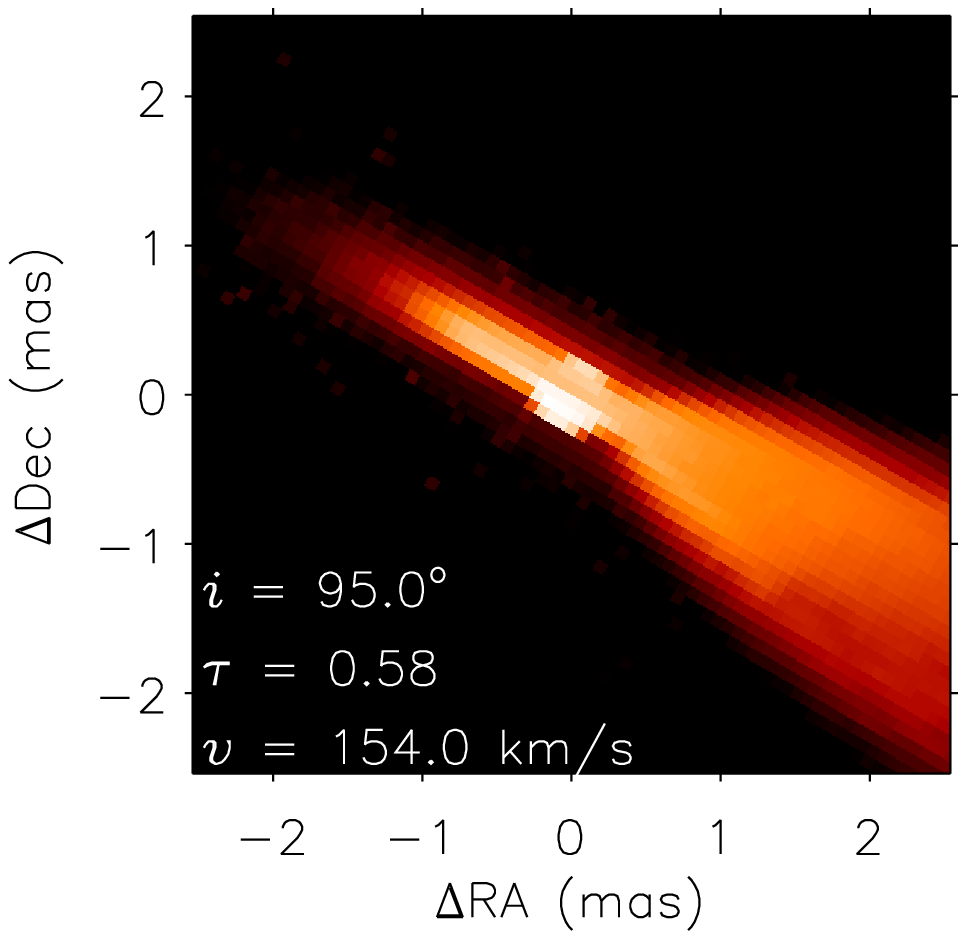}
      \caption{
      {\it Top left}: Density perturbation pattern of our best-fit model, as seen from above the disc. The line of sight towards the observer is marked for 5 $V/R$ phases. The VLTI/AMBER observations corresponds to $\tau=0.57$. The two circles mark the position on the disc for $\varpi = 2\,R_\mathrm{e}$ and $\varpi=5\,R_\mathrm{e}$.
      {\it Top center:} Density perturbation pattern projected onto the plane of the sky, at the time of the VLTI/AMBER observations. The solid line indicates the position of the stellar rotation axis. The dashed lines marks the orientations of the 3 baselines shown in Fig.~\ref{visib}.
      {\it Top right:} Continuum synthetic image at $2.16\,\mu\rm m$. Apparent from the plot is the brighter southern hemisphere of the star, which is little affected by the presence of the geometrically thin disc.
      {\it Bottom panels}: Synthetic images for three wavelengths across Br$\gamma$, as indicated              }
         \label{images}
   \end{figure*}

Perhaps the strongest evidence in support of the global oscillation scenario comes from the very good fit of the interferometric data.
The formal fit of the observed visibilities and phases is shown in Fig.~\ref{visib} for three baselines (marked as A, B and C, see Fig.~\ref{images} \nada{for their orientation with respect to the disc}).
In the figure, the left column shows the phases and the middle column the square visibilities for each baseline. Because of the poor absolute calibration of our dataset, both observables were normalized to the value obtained in the continuum. The difference in amplitude of the different profiles are explained by both the baseline lengths ($93$m, $53$m and $130$m, respectively) and orientation ($43\degr$, $99\degr$ and $63\degr$ East from North, respectively). For instance, baseline A corresponds to an orientation close to the minimum elongation axis of the disc, which explains the small drop in visibility.
The quality of the fit is quantitatively expressed by the $\chi^2$  map shown in the right column.

For illustration purposes we have converted the three interferometric phases into an astrometric displacement for \nada{each spectral} bin across the line (top-right panel of Fig.~\ref{visib}). Open and \nada{filled} symbols indicate \nada{the} displacement along the disc \nada{at} minimum and maximum elongation axes, respectively. The typical S-shape astrometric signature of a rotating disc is clearly distorted: the redshifted part of the disc extends further away from the central star than the \nada{blueshifted} part ($300\,\mu$as and $150\,\mu$as respectively). This is a direct, illustrative proof of the presence of an asymmetry in the disc. 

The fits of the visibilities and phases were carried out in the following way. For a given disc model and inclination angle, we computed synthetic images for several wavelengths around the Br$\gamma$ line (see Fig.~\ref{images}) for several values of $\phi$, corresponding to different $V/R$ phases. This set of synthetic images were then rotated for several values of $\gamma$, corresponding to different orientations of the disc in the skyplane. Next, the visibilities and phases for each of the modified images were computed and compared to the observed visibilities and phases. A $\chi^2$ minimization procedure was carried out that produced the values of $\gamma$ and $\phi$ that best fitted the data\footnote{\nada{For the $\chi^2$ calculation only  the spectral channels in the range $[-700,+700]\,\rm km\,s^{-1}$ were used. The value of $\chi^2$ depends on the number of spectral channels, and in the continuum these channels are always well matched and therefore tend to artificially reduce the $\chi^2$.}}. Finally, we manually explored the parameter $d$ in the range $113-148\,$pc. 
However, we have always found that this had only a marginal impact on our fit and, therefore, we have chosen to use the distance determined from the 2D analysis, $d=126\,$pc.

This procedure was repeated for several disc models and inclination angles.
An analysis of the flux, polarization and $V/R$ properties was also carried out for the same models.
The model parameters of Table~\ref{table:model_parameters} correspond to the values that best reproduced all data analyzed, including the interferometry.

Our data analysis indicates that the interferometry is much more sensivite to some model parameters than others. The most sensitive parameter, and hence the \nada{best} constrained, is undoubtedly the disc orientation $\gamma=32\pm5\degr$. As discussed in Paper I, this value agrees well with previous polarimetric and interferometric determinations.
In particular, it  is compatible with the recently published value $\gamma=37\pm2\degr$, obtained in the $K^{\prime}$ band with the CHARA array \citep[][]{gie07}. \nada{Additionally, as shown in the middle-right panel of Fig.~\ref{visib}, the overall size of our models also} agrees with that measured by  \citet{gie07}. We note the CHARA and VLTI/AMBER estimates are independent because \citet{gie07} used absolutely calibrated broad band visibilities, whereas we have used differential phases and visibilities across a line. 

The $\chi^2$ map of Fig.~\ref{visib} shows that the parameter $\phi$ is not \nada{as} well constrained.
From the fit we \nada{find} that the position of the density wave at the time of the observation is $\phi(T=54\,081.3) = 90\pm25\degr$. The large uncertainty comes from the combined effects of the high inclination of the disc and the fact that the density perturbations \nada{are} spread over a large range in azimuth (see Fig.~\ref{images}).
Using Eq.~\ref{phi_T} we find \nada{that the position of the density wave at $\tau=0$, as given by the interferometry, is} $\phi^{\rm int}_0 = 295\pm25\degr$. Despite \nada{this} large uncertainly, this value is in \nada{agreement with} our best-fit value of $\phi_0=280\degr$ obtained primarily from the fit of the $V/R$ phases.
We emphasize that spectroscopic and interferometric determinations of the $\phi_0$ are completely independent, \nada{making} the above agreement relevant.

The analysis of the interferometric data for different values of $i$ (not shown) largely \nada{confirms} the value for the inclination angle obtained from the 2D model (Sect.~\ref{2dmodel}).
In addition, interferometry has allowed us to lift a historical degeneracy in the determination of $i$.
Recall from Fig.~\ref{geometry} that the inclination angle goes from 0 to $180\degr$. 
For $i<90\degr$ the top of the disc is facing the observer and the contrary is true for $i>90\degr$.

In the case of an axisymmetric disc, it is really difficult to distinguish \nada{between} $i>90\degr$ and $i<90\degr$. For an infinitely thin disc this distinction is impossible, since the disc aspect would be identical for the two cases. 
For a disc of non-zero geometrical thickness, radiative transfer effects create subtle differences in the aspect of each disc, but the differences are probably too small to be detected.
However, for a non-axisymmetric disc this distinction is possible, provided two conditions are fulfilled: 
1) spatially and spectrally resolved observations \nada{are} available in order to determine the rotation direction of the disc in the plane of the sky, \nada{so that the northward direction can be defined, and
2) the disc asymmetry is chiral, i.e., the disc is not superimposable on its mirror image.
We note that we adopt the usual convention that the angular momentum vector points north.}

\nada{Both of} the above conditions are fulfilled for $\zeta$ Tau.  
The VLTI/AMBER observations unambiguously \nada{determine} that the western side of the disc approaches the Earth. Thus, according to the definition of north and the value of $\gamma$ derived from the interferometry, we know that the north of the system is oriented $32\degr$ east of celestial north.
Condition 2 is also fulfilled, since the disc is highly non-axisymmetric because of the spiral density perturbation produced by the global $m=1$ oscillation. From the detailed model fitting of the interferometric observations, we find that the model with viewing angle $i=95\degr$ ($\chi^2=1.4$) fits the observations better than the corresponding model with $i=85\degr$ \nada{($\chi^2=2.4$, see also significant differences between the models in Fig.~\ref{visib})}. 
Therefore, the VLTI/AMBER data \nada{allow} us to determine that \emph{the bottom part of the disc of $\zeta$ Tau faces the Earth}. Note that this degeneracy is nearly impossible to be lifted with broadband interferometry data only, such as the CHARA data shown in the right-middle panel of Fig.~\ref{visib}.
We believe that this is the first time the degeneracy in the determination of $i$ was lifted for a Be star.

 %


\subsubsection{Geometrical Considerations}

The geometrical properties of $\zeta$ Tau derived above are illustrated in Fig.~\ref{images}. In the top left panel we plot the density perturbation pattern, as seen from above the disc. 
The figure marks the position of the observer's line of sight for 5 $V/R$ phases. The $\tau=0.57$  line of sight corresponds to the time of the interferometric observations.
The top center panel is a sketch of the disc density projected onto the skyplane at the time of the VLTI/AMBER observations.
The synthetic image for the IR continuum (top right panel) mimics the projected disc density as expected. The bottom panels of Fig.~\ref{images} show model images for different wavelengths across the Br$\gamma$ line. 
The combination of the disc asymmetries and the effects of rotation \nada{generates} a very complex brightness distribution.

From Fig.~\ref{images} we see that in December 2006 ($\tau=0.57$) the base of the spiral arm was behind the star, but the bulk of the material was located in the north-east.
This is consistent with the $V/R < 1$ value at the period of VLTI/AMBER observations (recall that the western side of the disc is approaching us).


\subsubsection{Polarization and Flux}

   \begin{figure}
   \centering
   \includegraphics[width=\columnwidth]{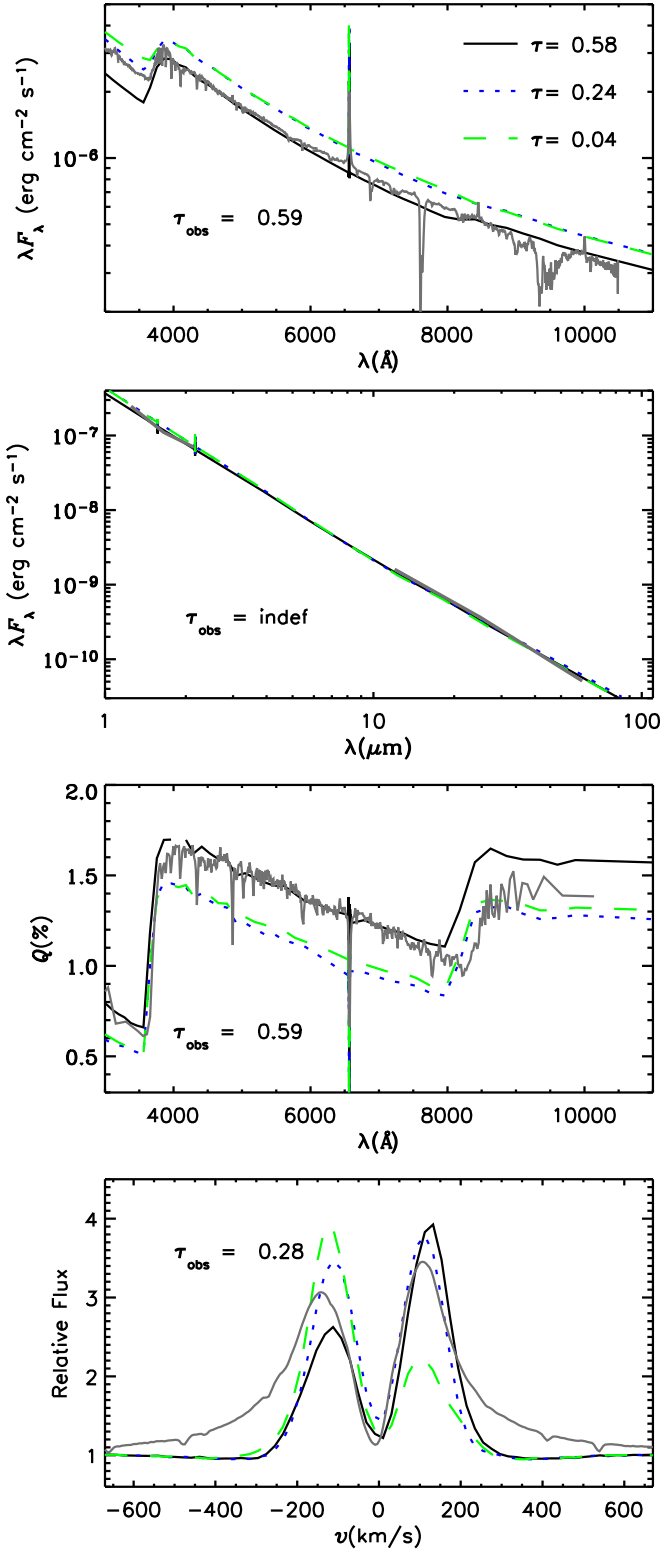}
      \caption{
      Emergent spectrum from $\zeta$ Tau. The observations are shown as the dark grey lines and the $V/R$ phase of each observation, $\tau_\mathrm{obs}$,  is indicated in each panel. Also plotted are the model predictions for three different $V/R$ phases, as indicated in the top panel.
    \emph{Top: } Visible SED \citep[data from][]{woo97}.
   \emph{Second from top:} IR SED (data from the 2MASS and IRAS  catalogs.
   \emph{Second from bottom: } Continuum polarization \citep[data from][]{woo97}.
   \emph{Bottom: } H$\alpha$ line profile (data from Paper I).
    As in Fig.~\ref{mosaic2d}, the light grey line corresponds to the unattenuated stellar SED
        }
    \label{mosaic3d}
   \end{figure}

    \begin{figure}
   \centering
   \includegraphics[width=\columnwidth]{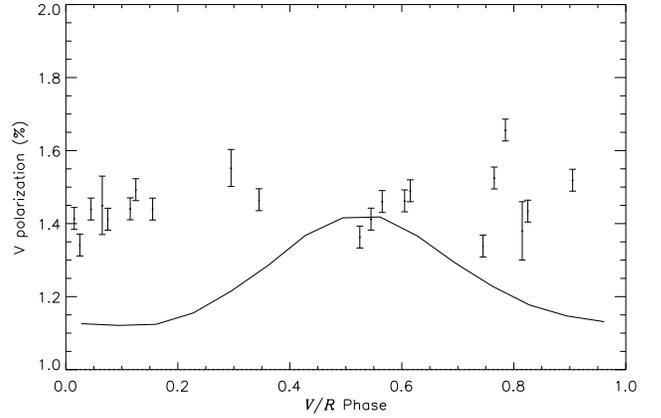}
      \caption{Temporal evolution of the polarization across the $V/R$ phase. The model polarization (line) is compared with the observations (points with error bars)
              }
         \label{pphase}
   \end{figure}

Besides the line $V/R$ variations, the model also predicts flux and polarization variations across one $V/R$ cycle.
This can be readily seen from Fig.~\ref{mosaic3d}, where we compare the observed SED, polarization spectrum and H$\alpha$ line profile with the model predictions for different phases.
The polarization variations are further illustrated in Fig.~\ref{pphase}.

What causes the variations in polarization is the occultation of part of the disc by the star in this nearly edge on system. Our models show that the polarization comes mainly from the region of the disc between $\varpi = 1$ to $\varpi \approx 5\,R_\mathrm{e}$. 
From the diagram in Fig.~\ref{images}, we see that the bulk of the density between those radii is in front of the star for $\tau\approx 0.5$. Since polarization is roughly proportional to the number of scatterers, the polarization is expected to be greatest at this phase, as in indeed the case (Fig.~\ref{pphase}). As the density perturbation pattern precesses, eventually the bulk of the material will move behind the star at $\tau\approx0$. The light scattered by this material will be partially absorbed by the star \nada{causing a decrease in the polarization}.

In Fig.~\ref{pphase} the predicted modulation of the polarization is compared to the observations \nada{(data from Paper I)}.
We have folded the data \nada{into} phase using the $V/R$ ephemerides of Table~\ref{table:model_parameters}. \nada{We plot only observations from 1992 to 2004.}
The polarization data seems to present a big challenge to the model. The model predicts large polarization variations of up to 40\%, and this is not observed. 
We will come back to this important point in Sect.~\ref{discussion}.

The flux excess in visible wavelengths comes from light emitted by the recombination of a free electron with a proton, \nada{commonly called bound-free radiation}. The recombination rates are roughly proportional to the square of the density and fall very rapidly with distance from the star. \nada{Thus the bulk of excess flux} in the visible comes from a very small region of the disc, out to approximately 2 stellar radii.
If we consider this small region, the density enhancement due to the global oscillations is behind the star for $\tau\approx0.5$ and in front of it for $\tau\approx0$, just the opposite situation described above for the polarization (Fig.~\ref{images}). This explains why the flux is smaller for $\tau=0.58$ in Fig.~\ref{mosaic3d}. 
The model does not predict important variations of the IR fluxes (second panel of Fig.~\ref{mosaic3d}) because in this case \nada{the} emission comes from a much larger fraction of the disc and occultation by the star \nada{has} a smaller effect.

Unfortunately, to our knowledge no systematic photometric monitoring was carried out for $\zeta$ Tau after the onset of the $V/R$ variations in 1992, so our predictions for the brightness variations cannot be tested at the moment.

We end this section \nada{with a discussion about} the role of the parameter $\Sigma_0$ in our 3D analysis.
When the global oscillations are considered,  $\Sigma_0$ represents the azimuthally averaged density scale of the disc.
\nada{We found from our  3D analysis a value of  $\Sigma_0$ 25\% larger than that obtained from the 2D analysis (see Sect.~\ref{2dmodel}).}

\section{Discussion \label{discussion}}

Our analysis of the data using the global oscillation scenario proposed by \citet{oka97} and \citet{pap92} shows that this model has both strong and weak points.
\nada{We have been able to} reproduce many observational characteristics of $\zeta$ Tau. Of particular relevance is the overall good fit of the $V/R$ cycles of H$\alpha$ and Br$\gamma$ lines and the excellent fit  to the VLTI/AMBER observations.
This favorable outcome of the analysis implies that the general morphology of the asymmetries in the disc of $\zeta$ Tau are correctly reproduced by the model, at least in the regions of the disc where most 
of the H$\alpha$ and Br$\gamma$ radiation \nada{originates}.
In addition, the fit of the observed phase difference between the $V/R$ cycles of H$\alpha$ and Br$\gamma$ strongly supports the theoretical prediction that the density wave has a helical shape.

However, two weak points of the model stand out from our analysis. Our results did not reproduce the phase of the $V/R$ cycle of the Br15 line, even though the amplitude was reproduced. Also, the model predicts large variations in polarization across one $V/R$ \nada{cycle that} are not observed.
Both the Br15 line and the polarization share a common property: both are formed in the very inner part of the envelope within a few stellar radii.

\nada{Considering the above discussion, a pattern emerges. Observables that trace} the larger scale of the disc (H$\alpha$ and Br$\gamma$) are reproduced by the model, whereas observables that map the inner disc (Br15 and polarization) are not.
This suggests that the global oscillation model fails to reproduce the shape and/or amplitude of the density wave in the inner disc, but it does reproduce the general features of the outer disc.

We must bear in mind that there are at least two alternative explanations as to why the model does not fit the Br15 $V/R$ cycle. First, the site for line formation, as calculated by HDUST may be wrong.
\nada{If the size of the Br15 line emitting region were larger in our models, then} the phase difference would have been smaller.
Second, there are other physical processes that  affect the structure of the inner disc, and \nada{these may also contribute to this discrepancy. For example,} the non-isothermal effects on the disc surface density calculated by \citet{car08b}.  \nada{Also, the fact that the disc is unlikely to be} fed by a smooth mass transfer from the star, but instead by a series of outbursts \citep[e.g.][]{riv98}. \nada{In this case} the steady-state viscous decretion disc would not be a good representation for the inner disc.

The simultaneous constraints of \nada{both the} period and mode confinement,  together with
\nada{a sophisticated} model, allow us to place more stringent limits on the $V/R$ parameters than the previous analyzes \citep[e.g.,][]{vak98}. 
However, there are still uncertainties in some parameters. This is particularly true for the \nada{adopted hypothetical radiative force used} to adjust the oscillation period.
There is also uncertainty arising from the fact that we cannot constrain
each parameter independently. For example, a larger value of rotation parameter $k_2 (\Omega/\Omega_\mathrm{crit})^2$ makes the mode \nada{more} confined, whereas the mode becomes less confined for a larger value of viscosity parameter $\alpha$. Thus, a similar good fit can be obtained for different sets of $k_2 (\Omega/\Omega_\mathrm{crit})^2$ and $\alpha$. In order to remove such a degeneracy, a \nada{more precise determination of the basic stellar parameters must be made.}



\section{Conclusions \label{conclusions}}

\nada{We have successfully completed a very detailed modeling of the Be star $\zeta$ Tau, using a very large set of observations that were reported in the first paper of this series.
$\zeta$ Tau provides a unique opportunity for testing theoretical ideas about disc formation and structure and about the origin of $V/R$ variations.}
First, the disc is known to have its global properties constant since at least 1992. 
Second, the presence of the binary \nada{companion} makes it reasonable to assume that the disc is truncated at the tidal radius of the system. Finally, the inclination angle of the disc is very well constrained.

We have assumed that the unperturbed disc structure is that of a steady-state (i.e., constant decretion rate) viscous Keplerian disc. We further assume that the disc is in vertical hydrostatic \nada{equilibrium, and to} describe the density perturbations we have adopted the global oscillation model of \citet{oka91} and \citet{pap92}.

We employed the computer code HDUST of \citet{car06a} to solve the coupled problems of radiative transfer, radiative equilibrium and statistical equilibrium and \nada{we fit, in a simultaneous and self-consistent way, all of the available observations, including the recent VLTI/AMBER observations} made in December, 2006.

The main conclusions of our data analysis are
 \begin{itemize}
\item We have obtained an excellent fit of the global, time averaged properties of the \nada{disc}.
From the theoretical point of view, this represents strong evidence that the viscous decretion disc scenario is the mechanism responsible for the formation of the CS disc of $\zeta$ Tau.

\item The \nada{agreement between our predictions and} the VLTI/AMBER observations and the acceptable fit of the $V/R$ cycle of the H$\alpha$ and Br$\gamma$ lines represent the first fairly comprehensive quantitative test of the global oscillation scenario. By fitting the spatially resolved interferometric observations we have shown that \nada{our model can reproduce the general features of the outer disc quite well. In particular, this demonstrates that the density wave is a spiral, as predicted}.

\item The model failed to reproduce the $V/R$ cycle of the Br15 line and predicted a large variation in the polarization \nada{during} a $V/R$ cycle, which is not observed. Since both the Br15 line and the polarization comes from the inner disc within a few stellar radii, this suggests that the global oscillation model does not predict the correct geometry in this region. However, other possibilities for the discrepancy between \nada{the model and the observations exist, including} non-isothermal effects in the inner disc structure and perturbations of the inner disc by outbursts.  A detailed analysis necessary to test both ideas will be left to future work.

\item With the aid of the VLTI/AMBER observations we have, for the first time, completely determined the geometrical properties of a Be star system. 
Non-spatially resolved determinations of the inclination angle (as, for instance, obtained from polarization studies) cannot distinguish between $i<90\degr$ (i.e., a disc whose northern face is towards the Earth) and  $i>90\degr$ (i.e., a disc whose southern face is towards the Earth).
Our detailed fit of the interferometric data allowed us to determine that the inclination angle of $\zeta$ Tau is $95\degr$, i.e., the disc is seen almost edge-on and \nada{its} southern face is towards us.
\end{itemize}

Recently, there has been important progress in the theory of global disc oscillations.
\citet{ogi08} found that the one-armed oscillation modes are naturally confined in discs larger than several tens of stellar radii, when the 3D structure of the oscillation mode is taken into account.
According to \citet{ogi08}, the vertical oscillation structure should not be neglected, because the vertical gravitational acceleration around an elliptical orbit of each gas particle in a disc perturbed by a one-armed oscillation mode excites an oscillatory vertical motion.

Given that 
$R_\mathrm{t} \lesssim 20\,R_\mathrm{e}$, 
the one-armed modes are unlikely to be confined in the $\zeta$ Tau disc even if this effect is taken into account. The amplitude of the oscillation, however, is expected to be significantly influenced in the outer part of the disc \nada{where the major contribution to the Balmer emission arises. In} the next paper of this series, we will apply the treatment by \citet{ogi08} to the $\zeta$ Tau disc, in order to study the effect of the vertical oscillations on this \nada{star and see if the model parameters can be well constrained and if these parameters agree with the values obtained by this study}.

\begin{acknowledgements}
\nada{The authors are very grateful to the referee, Dr. Carol Jones, for her useful comments and detailed review of the manuscript.} 
This work was supported by FAPESP grant 04/07707-3 to A.C.C., NSF  grant AST-0307686 to the University of Toledo (J.E.B.) and JSPS grant 20540236 to A.T.O.
Part of the calculations were carried out on the Itautec High Performance Cluster 
of the IAG/USP, whose purchase was made 
possible by the Brazilian agency FAPESP (grant 2004/08851-0).
A.C.C. is grateful for the help of the technical computing staff of the Astronomy Departament of the IAG/USP.
\end{acknowledgements}

\end{document}